\documentclass[final]{cvpr}

\usepackage{times}
\usepackage{epsfig}
\usepackage{graphicx}
\usepackage{amsmath}
\usepackage{amssymb}
\usepackage{dblfloatfix}
\usepackage[table,x11names]{xcolor}
\usepackage[normalem]{ulem}

\usepackage[pagebackref=true,breaklinks=true,colorlinks,bookmarks=false]{hyperref}

\def\cvprPaperID{****} 
\def\confYear{CVPR 2021}
\definecolor{grayhighlight}{RGB}{213,229,255}

\newsavebox\CBox
\def\textBF#1{\sbox\CBox{#1}\resizebox{\wd\CBox}{\ht\CBox}{\textbf{#1}}}

\begin{document}

\title{Fast and Accurate Camera Scene Detection on Smartphones}

\author{Angeline Pouget
\and Sidharth Ramesh
\and Maximilian Giang
\and Ramithan Chandrapalan
\and Toni Tanner
\and Moritz Prussing
\and Radu Timofte
\and Andrey Ignatov \vspace{2mm}
\and ETH Zurich, Switzerland}

\maketitle

\begin{abstract}

AI-powered automatic camera scene detection mode is nowadays available in nearly any modern smartphone, though the problem of accurate scene prediction has not yet been addressed by the research community. This paper for the first time carefully defines this problem and proposes a novel Camera Scene Detection Dataset (CamSDD) containing more than 11K manually crawled images belonging to 30 different scene categories. We propose an efficient and NPU-friendly CNN model for this task that demonstrates a top-3 accuracy of 99.5\% on this dataset and achieves more than 200 FPS on the recent mobile SoCs. An additional in-the-wild evaluation of the obtained solution is performed to analyze its performance and limitation in the real-world scenarios. The dataset and pre-trained models used in this paper are available on the project website.


\end{abstract}
{\let\thefootnote\relax\footnotetext{%
\hspace{-5mm}$^*$ Andrey Ignatov \textit{(andrey@vision.ee.ethz.ch)} and Radu Timofte \textit{(radu.timofte@vision.ee.ethz.ch)} are the main contacts. The dataset and the models presented in this paper are available on the project website: \\ \url{https://people.ee.ethz.ch/~ihnatova/camsdd.html}
}}

\section{Introduction}

\begin{figure*}[t!]
\resizebox{\linewidth}{!}
{
\large
\begin{tabular}{cccccc}
   \includegraphics[width=0.32\linewidth]{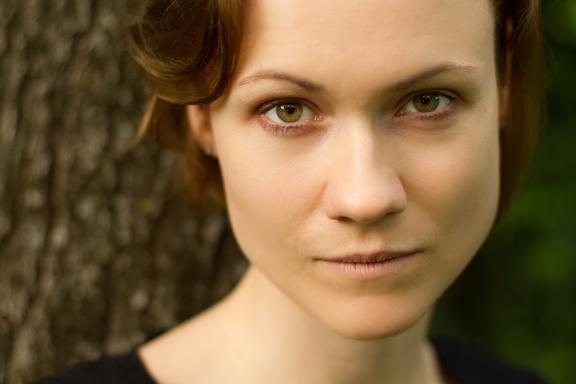}&
   \includegraphics[width=0.32\linewidth]{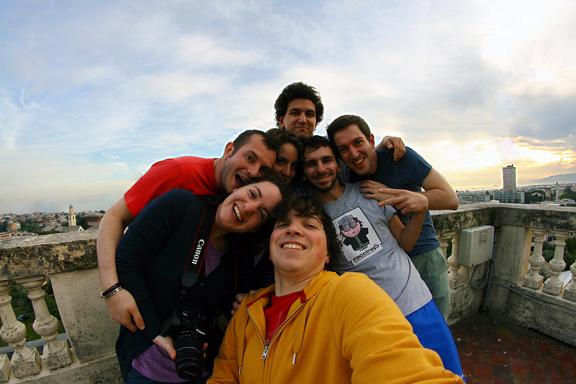}&
   \includegraphics[width=0.32\linewidth]{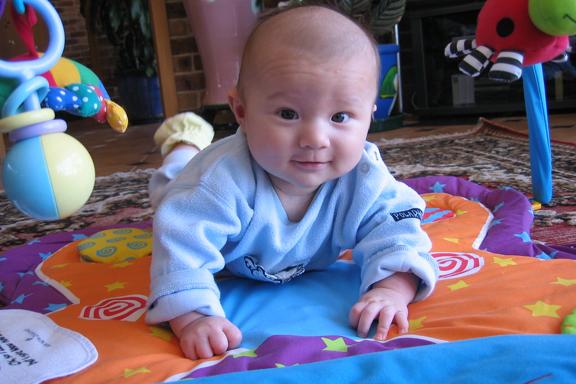}&
   \includegraphics[width=0.32\linewidth]{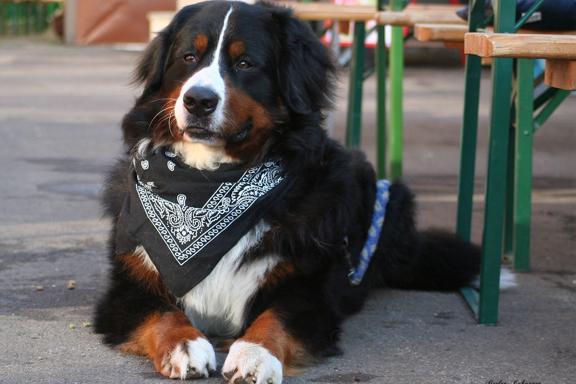}&
   \includegraphics[width=0.32\linewidth]{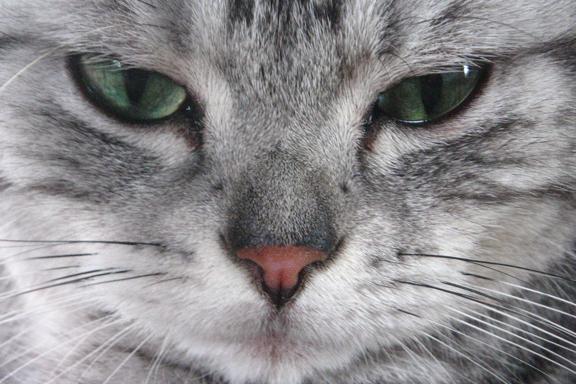}&
   \includegraphics[width=0.32\linewidth]{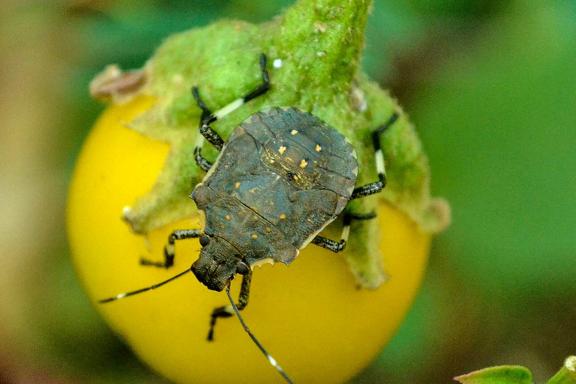}\\
Portrait & Group Portrait & Kids & Dog & Cat & Macro \\
   \includegraphics[width=0.32\linewidth]{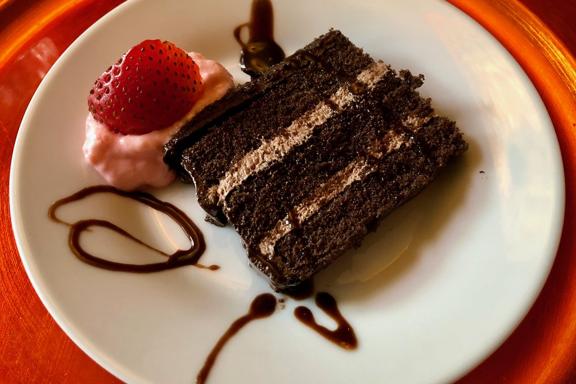}&
   \includegraphics[width=0.32\linewidth]{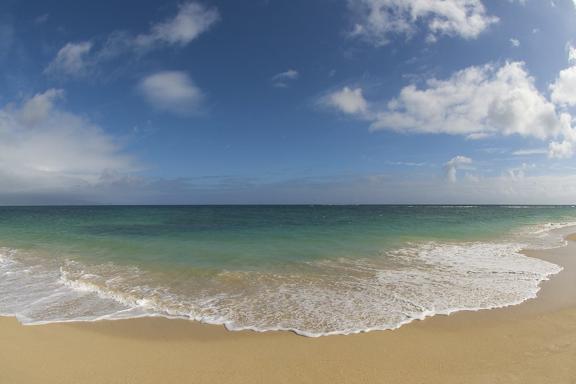}&
   \includegraphics[width=0.32\linewidth]{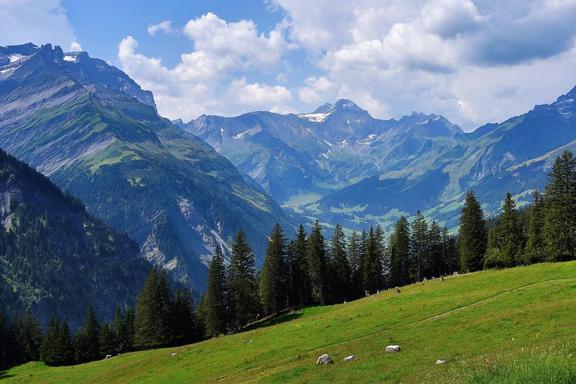}&
   \includegraphics[width=0.32\linewidth]{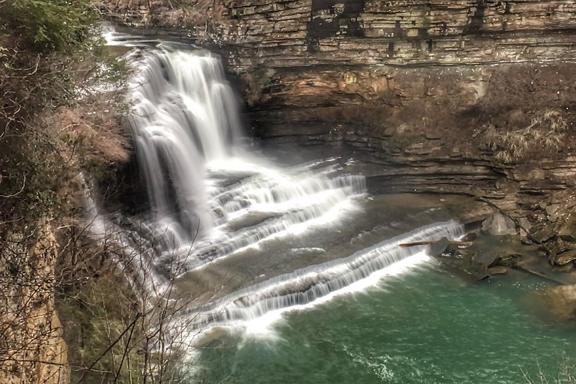}&
   \includegraphics[width=0.32\linewidth]{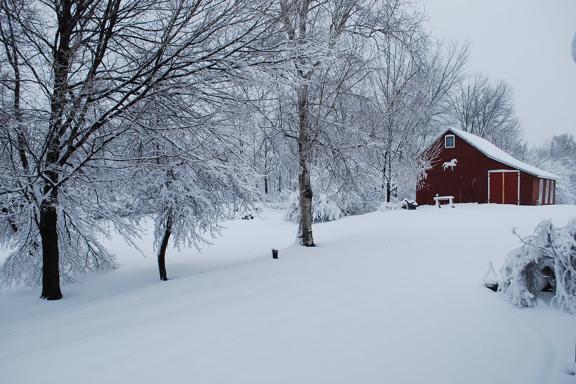}&
   \includegraphics[width=0.32\linewidth]{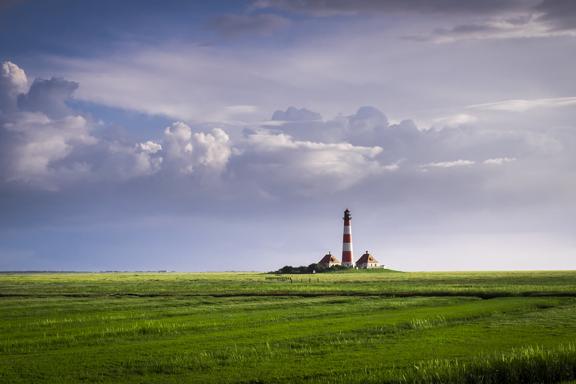}\\
Gourmet & Beach & Mountains & Waterfall & Snow & Landscape \\
   \includegraphics[width=0.32\linewidth]{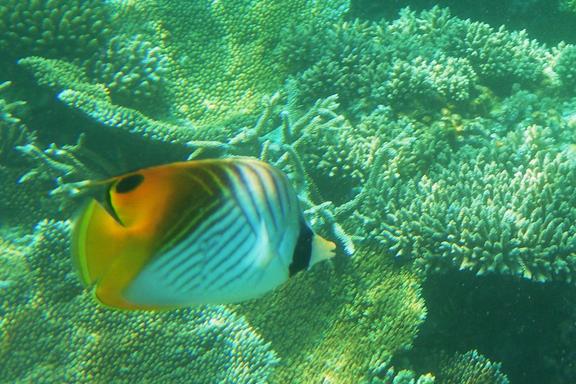}&
   \includegraphics[width=0.32\linewidth]{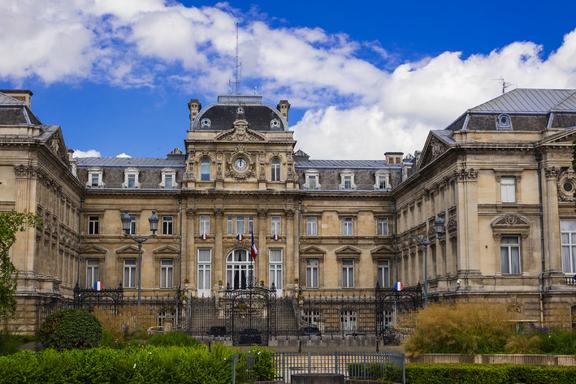}&
   \includegraphics[width=0.32\linewidth]{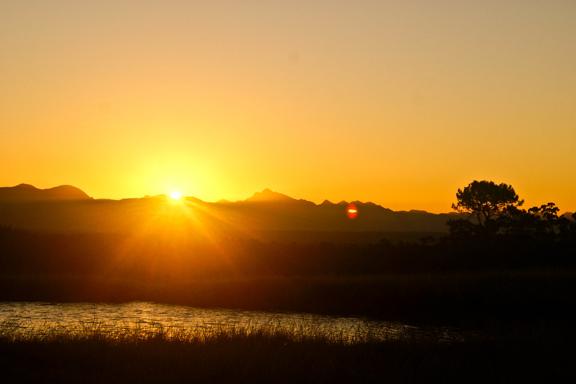}&
   \includegraphics[width=0.32\linewidth]{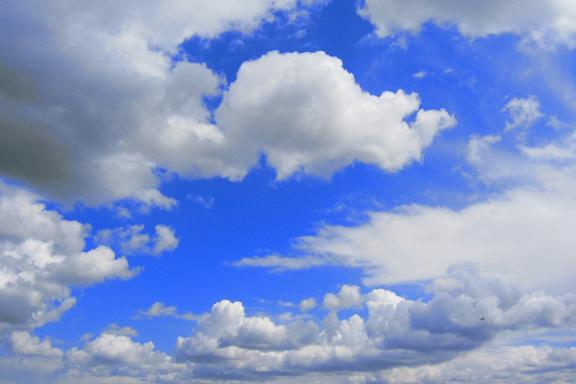}&
   \includegraphics[width=0.32\linewidth]{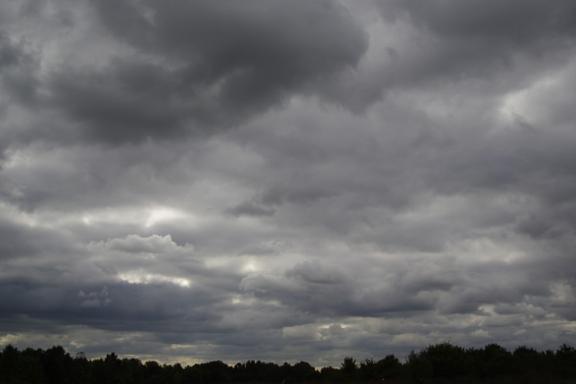}&
   \includegraphics[width=0.32\linewidth]{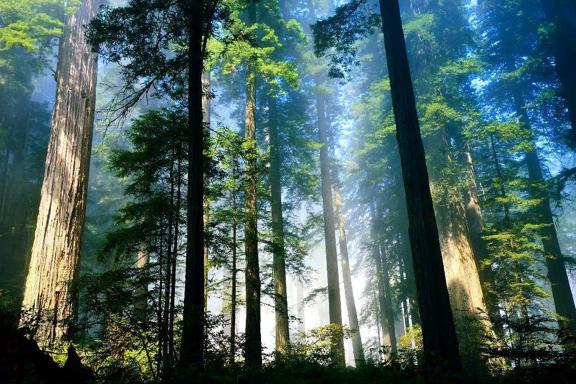}\\
Underwater & Architecture & Sunrise \& Sunset & Blue Sky & Overcast & Greenery \\
   \includegraphics[width=0.32\linewidth]{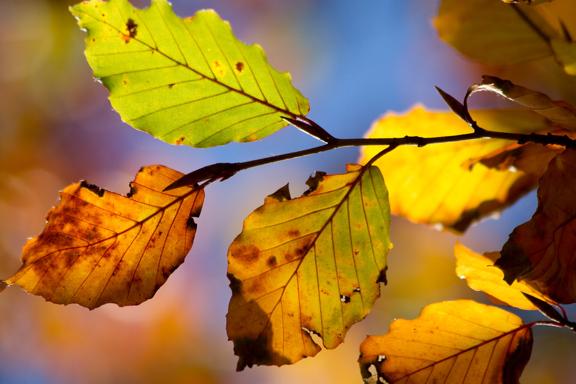}&
   \includegraphics[width=0.32\linewidth]{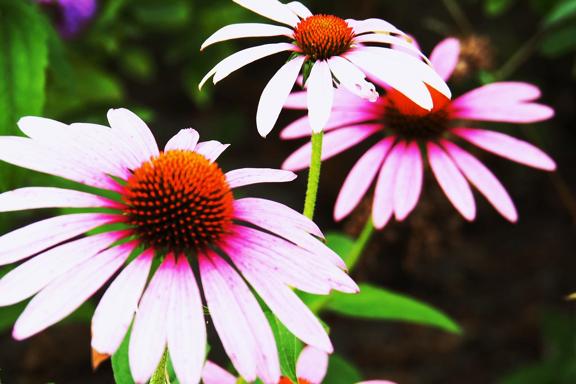}&
   \includegraphics[width=0.32\linewidth]{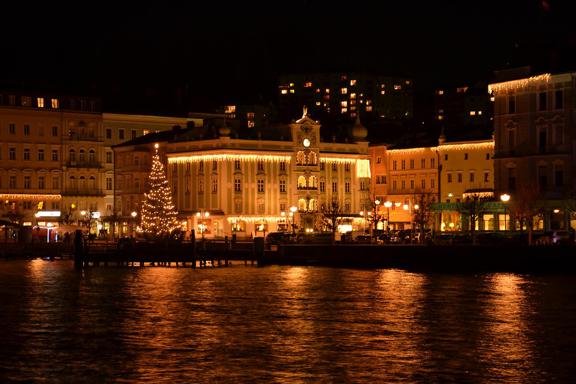}&
   \includegraphics[width=0.32\linewidth]{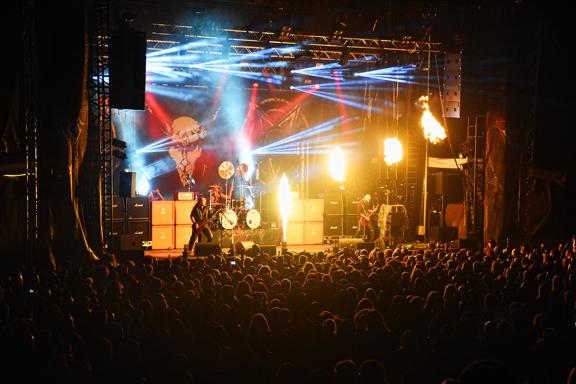}&
   \includegraphics[width=0.32\linewidth]{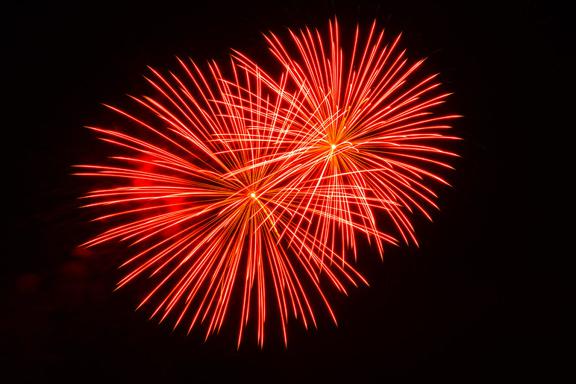}&
   \includegraphics[width=0.32\linewidth]{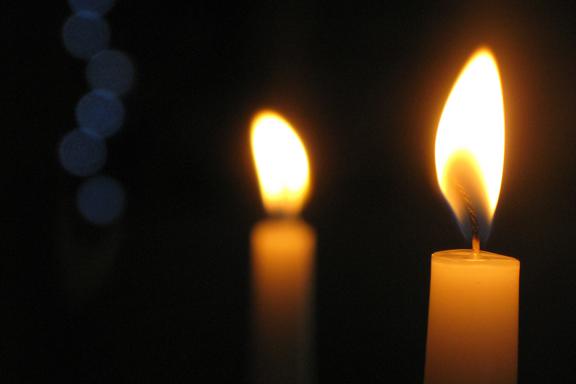}\\
Autumn Plants & Flowers & Night Shot & Stage & Fireworks & Candlelight \\
   \includegraphics[width=0.32\linewidth]{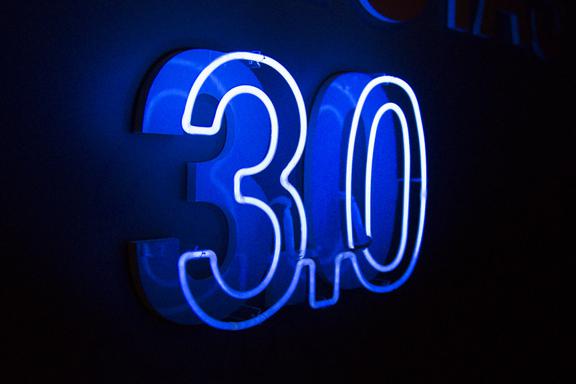}&
   \includegraphics[width=0.32\linewidth]{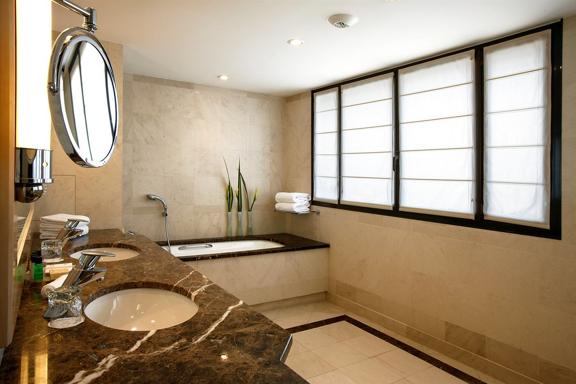}&
   \includegraphics[width=0.32\linewidth]{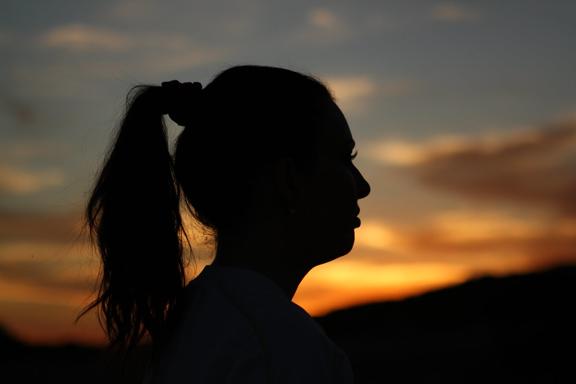}&
   \includegraphics[width=0.32\linewidth]{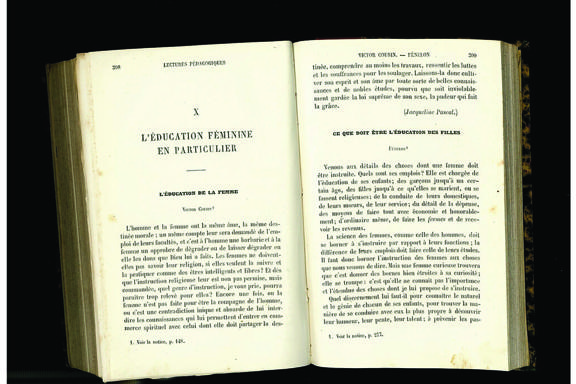}&
   \includegraphics[width=0.32\linewidth]{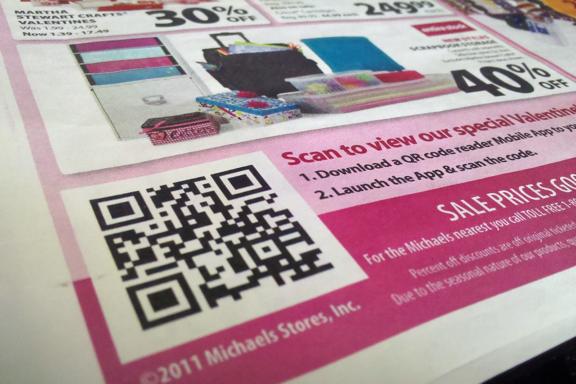}&
   \includegraphics[width=0.32\linewidth]{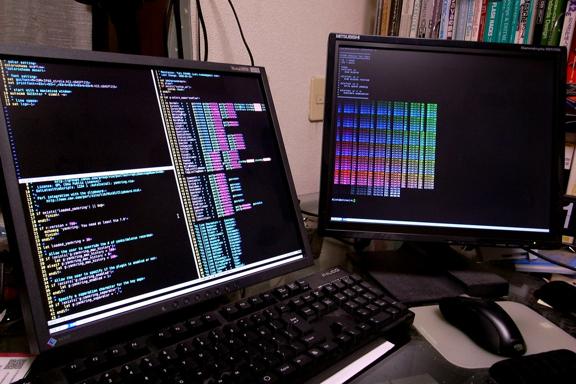}\\
Neon Lights & Indoor & Backlight & Document & QR Code & Monitor Screen \\
\end{tabular}
}
\vspace{2.0mm}
\caption{Visualization of the 30 Camera Scene Detection Dataset (CamSDD) categories.}
\vspace{0.0mm}
\label{fig:dataset}
\end{figure*}

Camera scene detection is one of the most popular computer vision problems related to mobile devices. Nokia N90 released in 2005 was the world's first smartphone with a manual camera scene selection option containing five categories (close-up, portrait, landscape, sport, night) and different lighting  conditions (sunny, cloudy, incandescent, fluorescent)~\cite{gsmarena2021nokian90}. Notably, it was also able to select the most appropriate scene automatically, though only basic algorithms were used for this and the result was not always flawless. Since then, this became a standard functionality for the majority of camera phones: it is applied to accurately adjust the photo processing parameters and camera settings such as exposure time, ISO sensitivity or white balancing to get the best image quality for various different scenes. For instance, certain situations require a high shutter speed to avoid the picture being blurry. A good example of this are pictures of animals, sport events or even kids. A modified tone mapping function is often needed for portrait photos to get a natural skin color, while special ISO sensitivity levels are necessary for low-light and night photography. An appropriate white balancing method should be used for indoor photos with artificial lighting so that the resulting images have correct colors. Finally, macro and portrait photos are often shot using bokeh mode~\cite{ignatov2020rendering} that should be enabled automatically for these scenes. Therefore, the importance of the camera scene detection task cannot be underestimated as it drastically affects the resulting image quality.

Using the automatic scene detection mode in smartphone cameras is very easy and convenient for the end user, but this poses the problem of making accurate predictions. The first scene classification methods were based on different heuristics and very simple machine learning-based algorithms as even the high-end mobile devices had at best a single-core 600 MHz Arm CPU at that time. The situation changed later when portable devices started to get powerful GPUs, NPUs and DSPs suitable for large and accurate deep learning models~\cite{ignatov2019ai,ignatov2018ai}. Since then, various AI-powered scene detection algorithms appeared in the majority of mobile devices from Huawei~\cite{aicamera2021huawei}, Samsung~\cite{aicamera2021samsung}, Xiaomi~\cite{aicamera2021xiaomi}, Asus~\cite{aicamera2021asus} and other vendors. However, since no available public datasets and models were available for this task, each manufacturer was designing its own solution that was often capable to recognize only a very limited number of classes.

To address the above problem, in this paper we present a novel large-scale \textit{CamSDD} dataset containing more than 11 thousand images and consisting of the 30 most important scene categories selected by analyzing the existing commercial solutions. We propose several efficient MobileNet-based models for the considered task that are able to achieve a top-1 / top-3 accuracy of more than 94\% and 99\%, respectively, and can run at over 200 FPS on modern smartphones. Finally, we perform a thorough performance evaluation of the proposed solution on smartphones in-the-wild and test its predictions for numerous real-world scenes.

The rest of the paper is arranged as follows. Section~\ref{sec:review} reviews the existing works related to image classification and efficient deep learning-based models for mobile devices. Section~\ref{sec:camsdd} introduces the CamSDD dataset and provides the description of the 30 camera scene detection categories. Section~\ref{sec:architecture} presents the proposed model architecture and the training details. Section~\ref{sec:experiments} shows and analyzes quantitative results, in-the-wild performance and the runtime of the designed solution on several popular mobile platforms. Finally, Section~\ref{sec:conclusion} concludes the paper.

\section{Literature Review}
\label{sec:review}

\begin{table*}
\centering
\resizebox{1.0\linewidth}{!}
{
\begin{tabular}{c|c|c|c|c|c}
\,ID\, & Category & Description & \,ID\, & Category & Description \\
\hline
\hline
1  & Portrait           & Normal portrait photos with a single adult or child           & 16 & Blue Sky & Photos with a blue sky (at least 50\%) \\
2  & Group Portrait     & Group portrait photos with at least 2 people                  & 17 & Overcast / Cloudy Sky & Photos with a cloudy sky (at least 50\%) \\
3  & Kids / Infants     & Photos of kids or infants (less than 5-7 years old)           & 18 & Greenery / Green Plants & Photos containing trees, grass and general vegetation \\
4  & Dog                & Photos containing a dog                                       & 19 & Autumn Plants & Photos with colored autumn leaves \\
5  & Cat                & Photos containing a cat                                       & 20 & Flower & Photos of flowers \\
6  & Macro / Close-up   & Photos taken at very close distance (\(<\) 0.3m)              & 21 & Night Shot & Photos taken at night \\
7  & Food / Gourmet     & Photos with food                                              & 22 & Stage / Concert & Photos of concert / performance stages\\
8  & Beach              & Photos of the beach (with sand and / or water)                & 23 & Fireworks & Photos of fireworks \\
9  & Mountains          & Photos containing mountains                                   & 24 & Candlelight & The main illumination comes from candles or fire \\
10 & Waterfalls         & Photos containing waterfalls                                  & 25 & Neon Lights / Signs & Photos of neon signs or lights \\
11 & Snow               & Winter photos with snow                                       & 26 & Indoor & Indoor photos with mediocre or artificial lighting \\
12 & Landscape          & \, Landscape photos (w/o snow, beach, mountains, sunset) \,   & 27 & Backlight / Contre-jour & \, Photos taken against a bright light source / silhouettes \\
13 & Underwater         & Photos taken underwater with a smartphone                     & 28 & Text / Document & Photos of documents or text \\
14 & Architecture       & Photos containing buildings                                   & 29 & QR Code & Photos with QR codes \\
15 & Sunrise / Sunset   & Photo containing sunrise or sunset                            & 30 & Monitor Screen & Photos of computer, TV or smartphone screens \\
\end{tabular}
}
\vspace{1.2mm}
\caption{The description of the 30 camera scene detection categories from the CamSDD dataset.}
\label{tab:camsdd_categories}
\vspace{-0.2mm}
\end{table*}

\subsection{Datasets}

Choosing the appropriate database is crucial when developing any camera scene detection solution. Though there already exist several large image classification datasets, they all have significant limitations when it comes to the considered problem. The popular CIFAR-10~\cite{krizhevsky2009learning} database presents a large number of training examples for object recognition task, though offers only 10 classes and uses tiny 32$\times$32 pixel images. In~\cite{darlow2018cinic}, the extended CINIC-10 dataset was presented that combines the CIFAR-10 and the ImageNet~\cite{deng2009imagenet} databases and uses the same number of classes and image resolutions. In contrast to these two datasets, the Microsoft Coco~\cite{lin2014microsoft} object recognition and scene understanding database labels the images by using per-instance object segmentation. ADE20K from~\cite{zhou2017scene} is another dataset providing pixel-wise image annotations with 3 to 6 times larger number of object classes compared to the COCO. As our focus is not to process the contextual information but to categorize individual images as precisely as possible, these two datasets are unfortunately not perfectly suitable for the camera scene detection task.

The SUN dataset~\cite{xiao2010sun,patterson2012sun} combines attribute, object detection and semantic scene labeling, and is mainly limited to scenes in which humans interact. The Places dataset~\cite{zhou2014learning} offers an even larger and more diverse set of images for the scene recognition task and enables near-human semantic classification performance, though it does not contain the vast majority of important camera scene categories such as overcast or portrait photos. With around 1 million images per category, the LSUN~\cite{yu2015lsun} database exceeds the size of all previously mentioned datasets~--- this was made possible by using semi-automated labeling. Unfortunately, it contains only 10 scene and 20 object categories, the majority of which are also not suitable for our task.

\subsection{Image Classification Architectures}

Since our target is to create an image classifier that runs on smartphones, the model should meet the efficiency constraints imposed by mobile devices. MobileNets~\cite{howard2017mobilenets} were among the first models proposing both good accuracy and latency on mobile hardware. MobileNetV2~\cite{sandler2018mobilenetv2} aims to provide a simple network architecture suitable for mobile applications while being very memory efficient. It uses an inverted residual block with a linear bottleneck that allows to achieve both good accuracy and low memory footprint. The performance of this solution was further improved in~\cite{howard2019searching}, where the new MobileNetV3 architecture was obtained with the neural architecture search (NAS). This model was optimized to provide a good accuracy / latency trade-off, and is using hard-swish activations and a new lightweight decoder.

EfficientNet~\cite{tan2019efficientnet} is another architecture suitable for mobile use cases. It proposes a simple but highly efficient scaling method for convolutional networks by using the ``compound coefficient'' allowing to scale-up the baseline CNN to any target resource constraint. Despite the many advantages of this architecture and its top scores on the ImageNet dataset~\cite{imagenet2021benchmark}, its performance highly depends on the considered problem, and besides that it is not yet fully compatible with the Android Neural Networks API (NNAPI)~\cite{NNAPI2021}.

Similarly to the MobileNetV3, the MnasNet~\cite{tan2019mnasnet} architecture was also constructed using the neural architecture search approach with additional latency-driven optimizations. It introduces a factorized hierarchical search space to enable layer diversity while still finding a balance between flexibility and search space size. A similar approach was used in~\cite{xie2019exploring}, where the authors introduced the Randomly Wired Neural Networks which architecture was also optimized using NAS, and the obtained models were able to outperform many standard hand-designed architectures. A different network optimization option was proposed in~\cite{wang2018pelee}: instead of  focusing on depthwise separable convolutions, the PeleeNet model is using only conventional convolutional layers while showing better accuracy and smaller model size compared to the MobileNet-V2. Though this nework demonstrated better runtime on NVIDIA GPUs, no evidence for faster inference on mobile devices was, however, provided.

\subsection{Deep Transfer Learning}

Network-based deep transfer learning~\cite{tan2018survey} is an important tool in machine learning that tackles the problem of insufficient training data. The term denotes the reuse of a partial network that has been trained on data which is not part of, but similar in structure to the training data. This partial network serves as a feature extractor and its layers are usually frozen after the initial training. It has been shown that the features computed in higher layers of the network depend greatly on the specific dataset and problem which is why they are usually omitted for transfer learning~\cite{yosinski2014transferable}. In some cases, it can be advantageous to fine-tune the uppermost layers of this transferred network by unfreezing their weights during training. On top of the feature extractor, one or several fully connected, trainable layers are added that are task-specific. Their weights are initialized randomly and updated with the use of the training data. Hence this part of the network aims to replace the non-transferred part of the model backbone architecture.

\subsection{Running CNNs on Mobile Devices}

When it comes to the deployment of AI-based solutions on mobile devices, one needs to take care of the particularities of mobile NPUs and DSPs to design an efficient model. An extensive overview of smartphone AI acceleration hardware and its performance is provided in \cite{ignatov2019ai,ignatov2018ai}. According to the results reported in these papers, the latest mobile NPUs are already approaching the results of mid-range desktop GPUs released not long ago. However, there are still two major issues that prevent a straightforward deployment of neural networks on mobile devices: a restricted amount of RAM, and a limited and not always efficient support for many common deep learning layers and operators. These two problems make it impossible to process high resolution data with standard NN models, thus requiring a careful adaptation of each architecture to the restrictions of mobile AI hardware. Such optimizations can include network pruning and compression~\cite{chiang2020deploying,ignatov2020rendering,li2019learning,liu2019metapruning,obukhov2020t}, 16-bit / 8-bit~\cite{chiang2020deploying,jain2019trained,jacob2018quantization,yang2019quantization} and low-bit~\cite{cai2020zeroq,uhlich2019mixed,ignatov2020controlling,liu2018bi} quantization, device- or NPU-specific adaptations, platform-aware neural architecture search~\cite{howard2019searching,tan2019mnasnet,wu2019fbnet,wan2020fbnetv2}, \etc.

\section{Camera Scene Detection Dataset (CamSDD)}
\label{sec:camsdd}

\begin{figure}[t!]
\centering
\resizebox{0.8\linewidth}{!}
{
\includegraphics[width=1.0\linewidth]{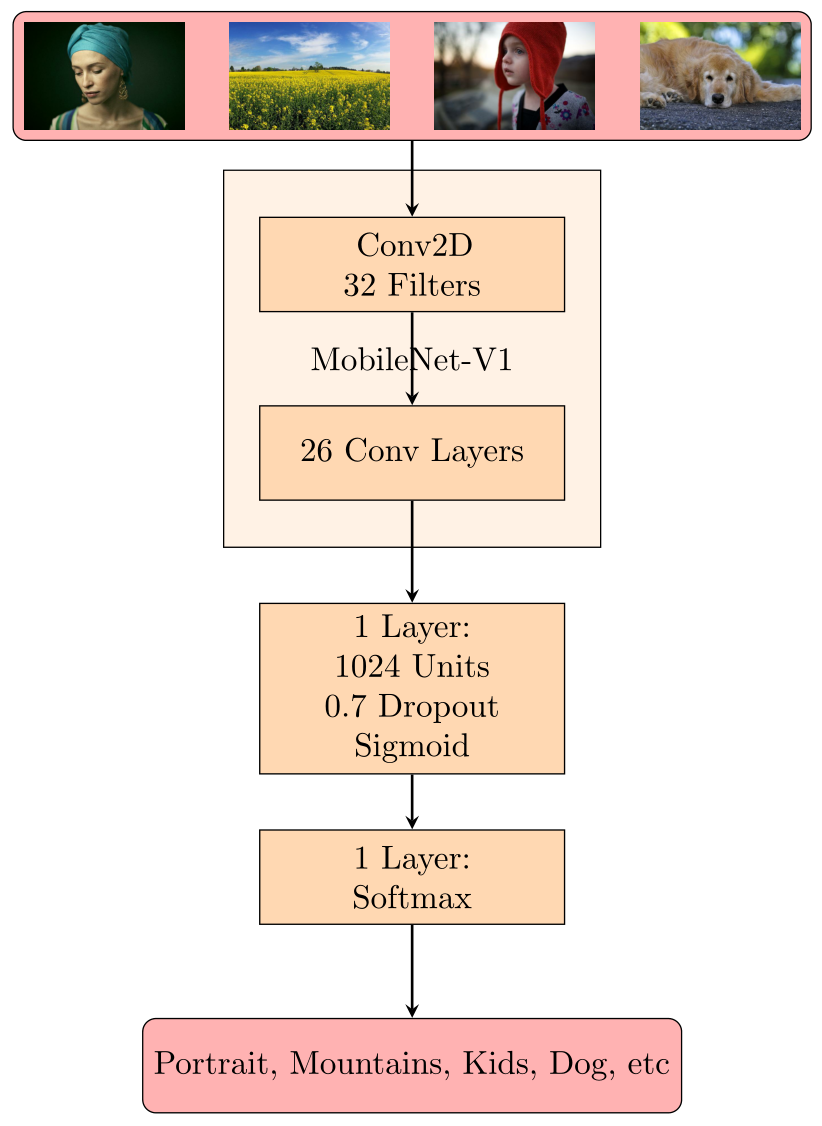}
}
\vspace{2mm}
\caption{\small{An overview of the MobileNet-V1 based model.}}
\vspace{-3.2mm}
\label{fig:architecture}
\end{figure}

When solving the camera scene detection problem, one of the most critical challenges is to get high-quality diverse data for training the model. Since no public datasets existed for this task, a new large-scale \textit{Camera Scene Detection Dataset (CamSDD)} containing more than 11K images and consisting of 30 different categories was collected first. The photos were crawled from \textit{Flickr}\footnote{\url{https://www.flickr.com/}} using the same setup as in~\cite{ignatov2018wespe}. All photos were inspected manually to remove monochrome and heavily edited pictures, images with distorted colors and watermarks, photos that are impossible for smartphone cameras (\eg, professional underwater or night shots), \etc. The dataset was designed to contain diverse images, therefore each scene category contains photos taken in different places, from different viewpoints and angles: \eg, the ``cat'' category does not only contain cat faces but also normal full-body pictures shot from different positions. This diversity is essential for training a model that is generalizable to different environments and shooting conditions. Each image from the CamSDD dataset belongs to only one scene category. The dataset was designed to be balanced, thus each category contains on average around 350 photos. After the images were collected, they were resized to 576$\times$384 px resolution as using larger photos will not bring any information that is vital for the considered classification problem. The description of all 30 categories is provided in Table~\ref{tab:camsdd_categories}, sample images from each category are demonstrated in Fig.~\ref{fig:dataset}. In the next sections, we will demonstrate that the size and the quality of the CamSDD dataset is sufficient to train a precise scene classification model.

\section{Method Description}
\label{sec:architecture}

\begin{figure*}[b!]
\centering
\resizebox{1.0\linewidth}{!}
{
\includegraphics[width=1.0\linewidth]{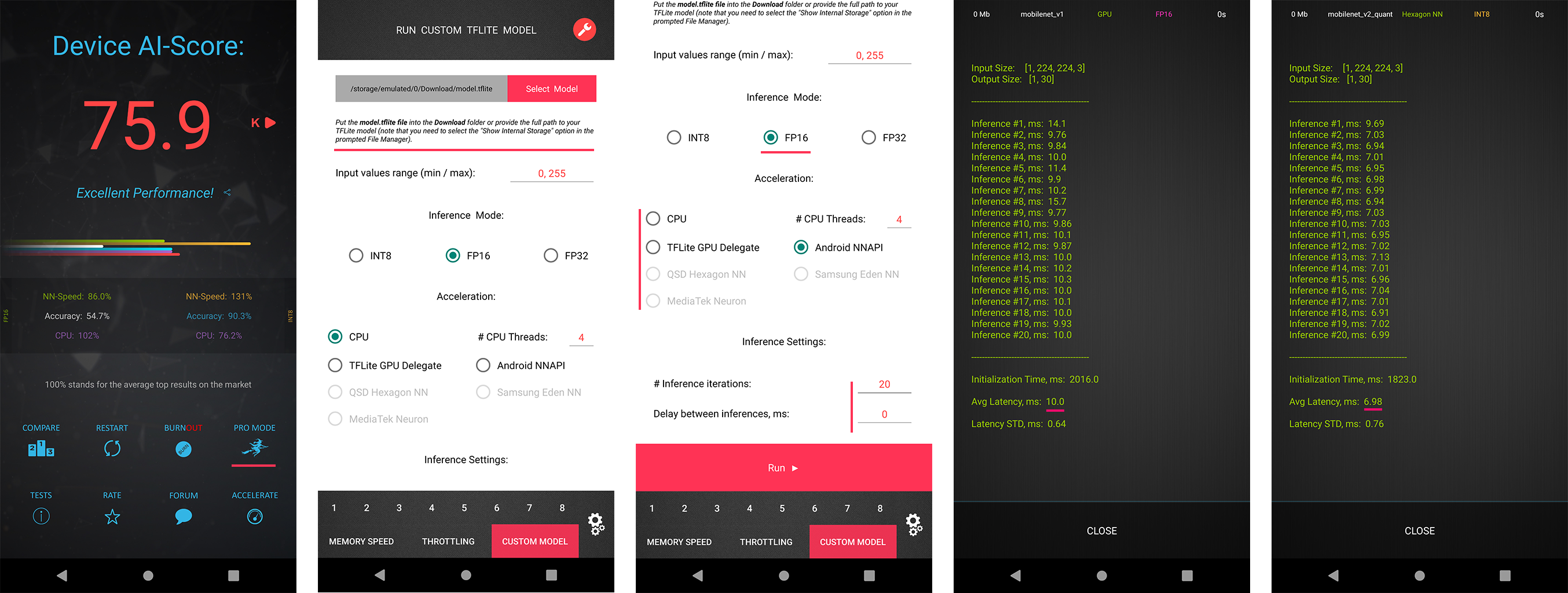}
}
\vspace{-2mm}
\caption{\small{Loading and running custom TensorFlow Lite models with AI Benchmark application. The currently supported acceleration options include Android NNAPI, TFLite GPU, Hexagon NN, Samsung Eden and MediaTek Neuron delegates as well as CPU inference through TFLite or XNNPACK backends. The latest app version can be downloaded at \url{https://ai-benchmark.com/download}}}
\vspace{-3.2mm}
\label{fig:ai_benchmark_custom}
\end{figure*}


This section provides a detailed overview and description of the designed solution and its main components.

\subsection{Feature Extraction}

\begin{table}
\centering
\resizebox{0.94\columnwidth}{!}
{
\begin{tabular}{l|cc}
Activation function & Top-1 Accuracy, \% & Top-3 Accuracy, \% \\
\hline
Sigmoid & 94.17 & 98.67 \\
ReLu & 93.33 & 98.17 \\
Tanh & 92.17 & 98.83 \\
SeLu & 92.00 & 98.17 \\
\end{tabular}
}
\vspace{3.2mm}
\caption{The accuracy of the MobileNet-V2 based model with different activation functions in the last fully-connected layer.}
\label{tab:activ_fun}
\vspace{-2.2mm}
\end{table}

Our proposed model architectures are built on the MobileNet-V1~\cite{howard2017mobilenets} and MobileNet-V2~\cite{sandler2018mobilenetv2} backbones. In general, MobileNets are based on depthwise separable convolutions except for the first layer which is fully convolutional. All layers are followed by batch normalization and use ReLU nonlinearity. There are two major reasons why these models are best suited to solve the challenge at hand. First, the MobileNet architectures are specifically tailored for mobile and resource-constrained environments. Due to the above mentioned depthwise convolutions, they perform a smaller number of operations and use less RAM while still retaining high accuracy on many image classification tasks. Due to these advantages, they are commonly used for a wide variety of applications, therefore NN HAL and NNAPI drivers of all vendors contain numerous low-level optimizations for the MobileNet architectures, which results in very efficient execution and small inference time on all mobile platforms.

We use all convolutional layers of these models with weights learned on the ImageNet dataset and omit only the fully-connected layers at the end. This has been shown to work best in contrast to replacing some of the convolutional layers as well. Intuitively, this observation makes sense since our main objective is to correctly predict the scene pictured in an image. This is also the main goal of the ImageNet Large Scale Visual Recognition Competition (ILSVRC)~\cite{imagenet2021competition}, an annual software contest run by the ImageNet project, though different image categories are used in this challenge. Due to this similarity in aims, the features of the input data that the MobileNets need to make an accurate final prediction, and the features that are crucial for our model are nearly the same, and thus retraining it on our data did not lead to better results on this task.

\subsection{Fully connected layers}

\begin{table}
\centering
\resizebox{1.0\columnwidth}{!}
{
\begin{tabular}{l|c|c|c}
Backbone & \,Model Size,\, & Top-1 & Top-3 \\
Architecture & MB & \, Accuracy, \% \, & \, Accuracy, \% \, \\
\hline
\hline
MobileNet-V1  & 208 & 92.67 & \textBF{99.50} \\
MobileNet-V2 & 73 & \textBF{94.17} & 98.67 \\
\hline
\textit{MobileNet-V1 Quantized} \,\, & 52 & 91.50 & 99.00 \\
\textit{MobileNet-V2 Quantized} \,\, & 19 & \textBF{94.17} & 98.67 \\
\hline
EfficientNet-B0 & 261 & 91.33 & 98.67 \\
MobileNet-V3 Small & 202 & 89.50 & 98.50 \\
MobileNet-V3 Large & 262 & 88.50 & 99.00 \\
Inception-ResNet-V2 & 359 & 86.00 & 97.00 \\
Inception-V3 & 284 & 85.50 & 96.33 \\
Xception & 472 & 86.33 & 98.17 \\
NASNetMobile & 220 & 66.00 & 84.67 \\
\end{tabular}
}
\vspace{2.6mm}
\caption{Top-1 and Top-3 classification accuracy of the proposed floating point and quantized MobileNet-V1/V2 based models. The results of the other architectures are provided for the reference.}
\label{tab:quant_results}
\vspace{-2.2mm}
\end{table}

\noindent\textBF{MobileNet-V1 Backbone.} On top of the the last convolutional layer of the MobileNet-V1, we placed a fully connected layer with $1024$ units and a dropout of $0.7$ to avoid overfitting. The activation in this layer is the \textit{Sigmoid} function which has worked best in comparison to other activation functions. The final output layer of the network uses the \textit{Softmax} activation to predict the probability of the input image belonging to any of the $30$ classes. An overview of the overall model structure is presented in Fig.~\ref{fig:architecture}.

\smallskip

\noindent\textBF{MobileNet-V2 Backbone.} A fully connected layer with $256$ units and the \textit{ReLU} activation function was placed on top of the last convolutional layer of the MobileNet-V2. It is followed by another fully connected layer with $1024$ units that uses \textit{ReLU} as well. The last fully connected layer has $512$ units with a dropout rate of $0.7$ to avoid overfitting. The activation in this last layer is the \textit{Sigmoid} function demonstrating the best top-1 accuracy compared to other activation functions such as \textit{SeLU}, \textit{ReLU}, or \textit{Tanh} as shown in Table~\ref{tab:activ_fun}. The final output layer of the network again uses the \textit{Softmax} activation to predict the actual scene category.

\subsection{Training Details}

\begin{table}[t!]
\centering
\resizebox{\linewidth}{!}
{
\begin{tabular}{l|cc|cc}
 & \multicolumn{2}{c|}{ MobileNet-V1 } & \multicolumn{2}{c}{MobileNet-V2} \\
Mobile SoC & \small{\, FP16, fps\,} & \small{\, INT8, fps \,} & \small{\, FP16, fps \,} & \small{\, INT8, fps \,} \\
\hline
\hline
Dimensity 1000+ \, & \textBF{220} & \textBF{222} & \textBF{224} & \textBF{233} \\
Dimensity 800 & 155 & 203 & 159 & 209 \\
Helio P90 & 43 & 52 & 48 & 46 \\
\hline
Snapdragon 888 \, & 136 & \small{72$^*$} & 126 & \small{76$^*$} \\
Snapdragon 855 & 100 & 113 & 85 & 143 \\
Snapdragon 845 & 75 & 65 & 79 & 88 \\
\hline
Exynos 2100 & 88 & 85 & 68 & 101 \\
Exynos 990 & 49 & 71 & 48 & 79 \\
Exynos 9820 & 59 & 52 & 56 & 56 \\
\hline
Kirin 990 5G & 50 & \small{81$^*$} & 132 & \small{86$^*$} \\
Kirin 980 & 33 & \small{74$^*$} & 42 & \small{78$^*$} \\
\end{tabular}
}
\vspace{2.6mm}
\caption{\small{The speed of the proposed solutions on several popular mobile SoCs. The runtime was measured with the AI Benchmark app using the fastest acceleration option for each device. $^*$~These results were obtained on CPU (4 threads) as the device was unable to parse the corresponding quantized TensorFlow Lite models.}}
\label{tab:runtime_results}
\end{table}

The models were implemented in TensorFlow~\cite{abadi2016tensorflow} and trained with a batch size of $20$ using the Adam optimizer~\cite{kingma2014adam}. The initial rate was set to $10^{-4}$ with an exponential decay of $0.1$ every $3$ epochs. In general, the performance of the model saturated after less than $15$ epochs of training. In case of the MobileNet-V2 based network, its convolutional layers were unfreezed after the initial training, and the entire model was additionally fine-tuned for few epochs with a learning rate of $10^{-5}$.

\begin{figure*}[t!]
\resizebox{\linewidth}{!}
{
\begin{tabular}{cccccc}
   \includegraphics[width=0.32\linewidth]{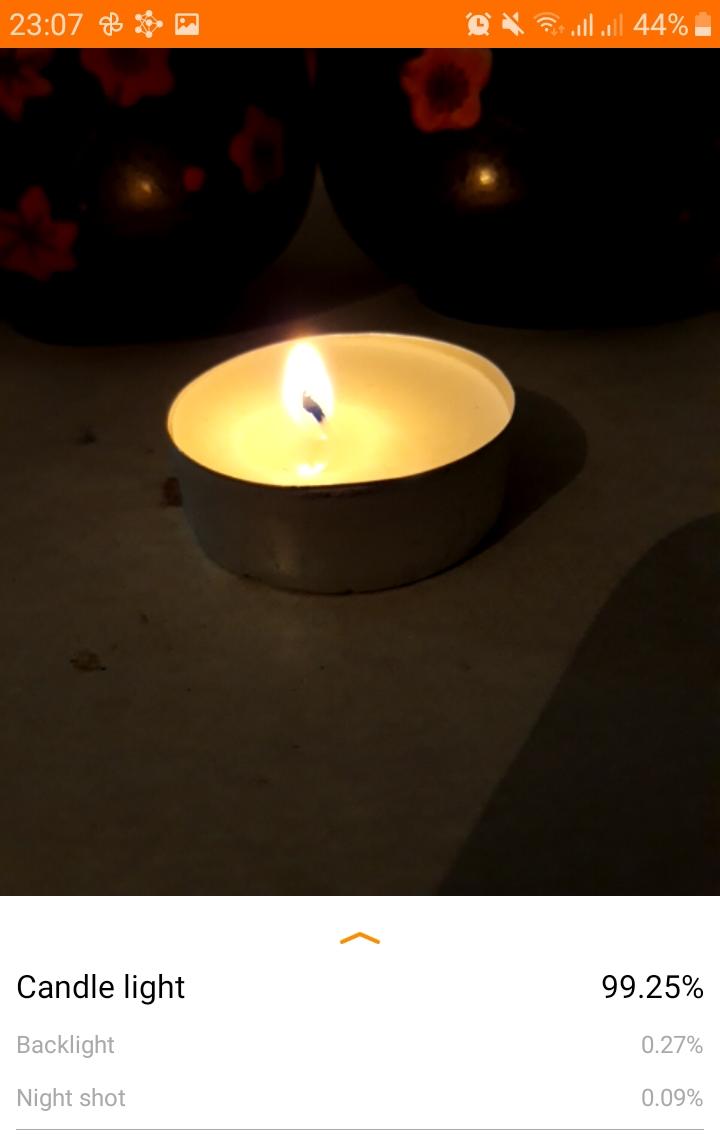}&
   \includegraphics[width=0.32\linewidth]{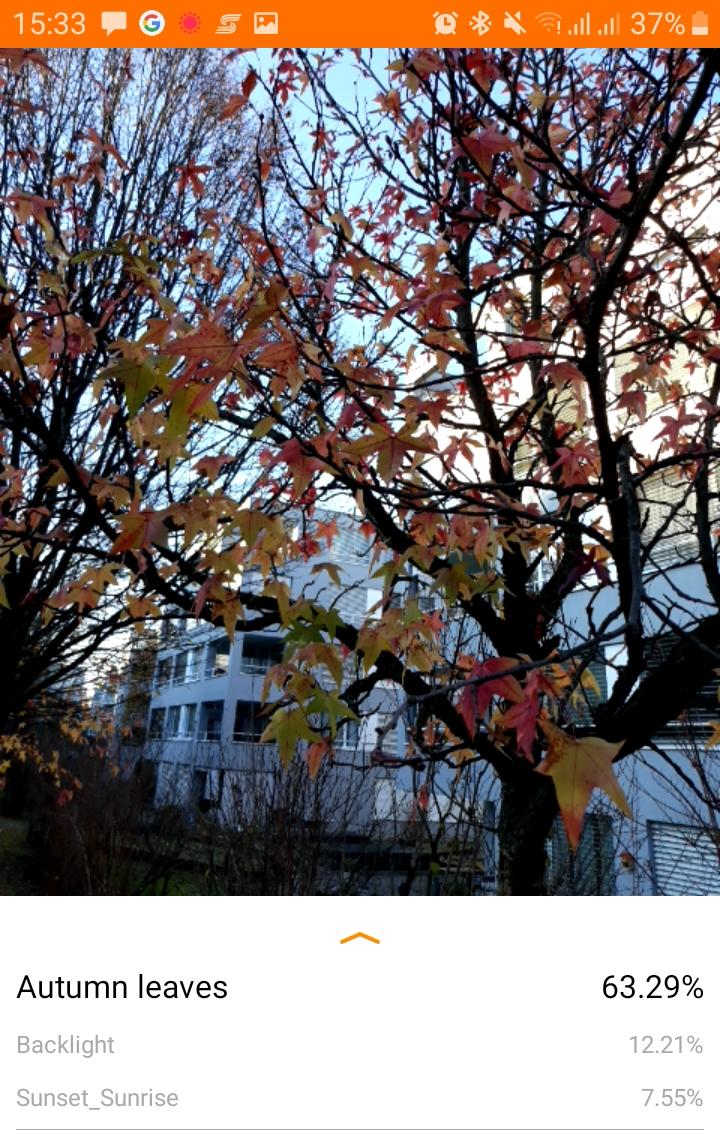}&
   \includegraphics[width=0.32\linewidth]{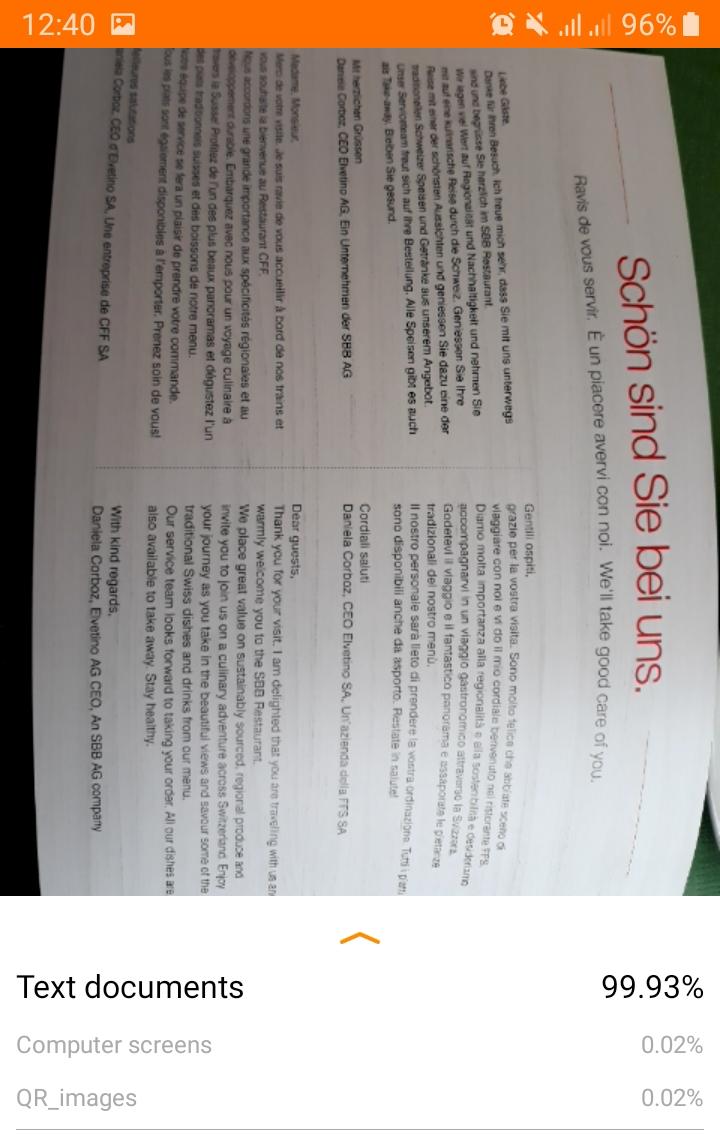}&
   \includegraphics[width=0.32\linewidth]{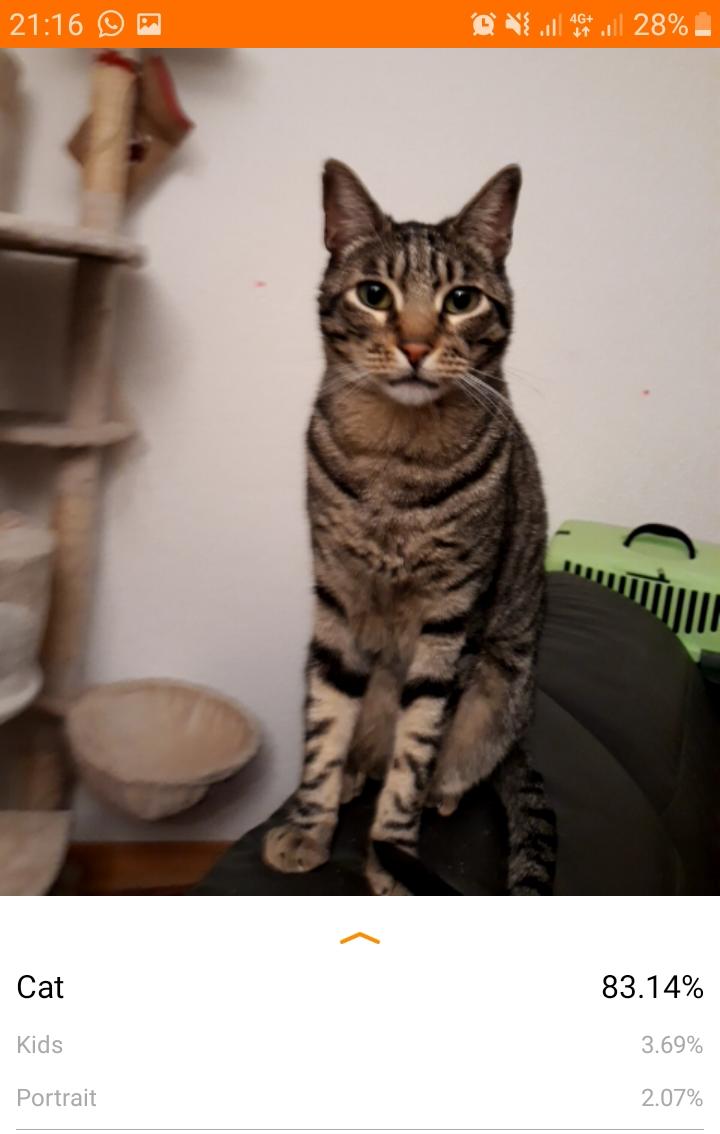}&
   \includegraphics[width=0.32\linewidth]{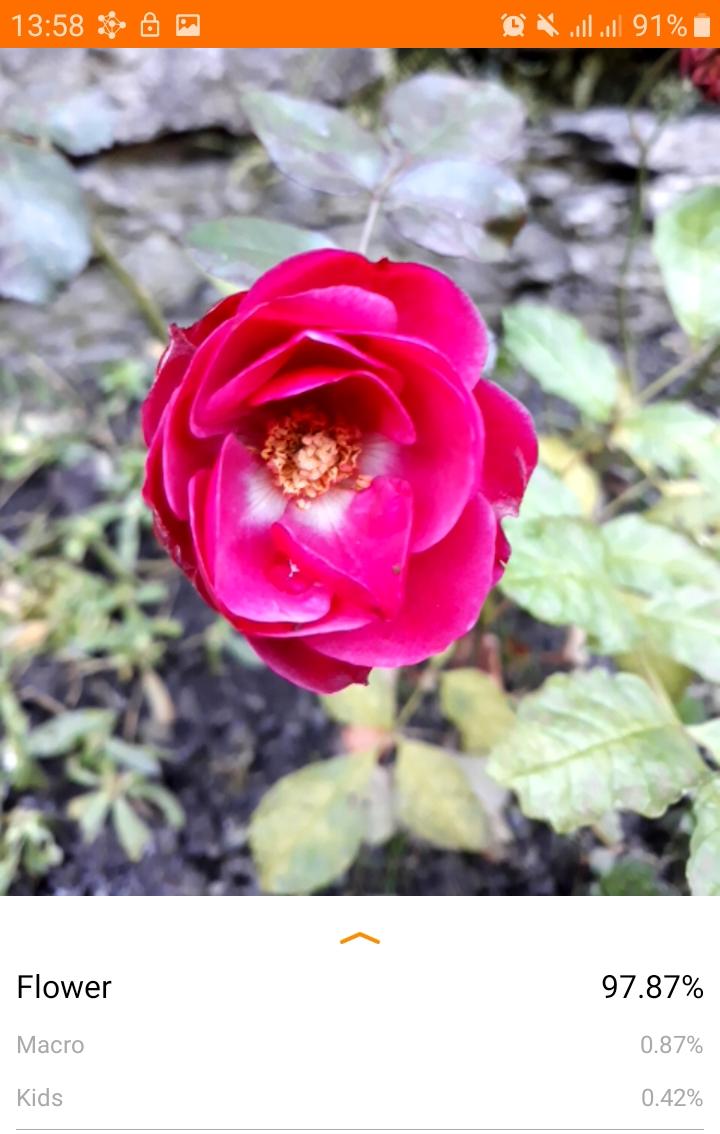}&
   \includegraphics[width=0.32\linewidth]{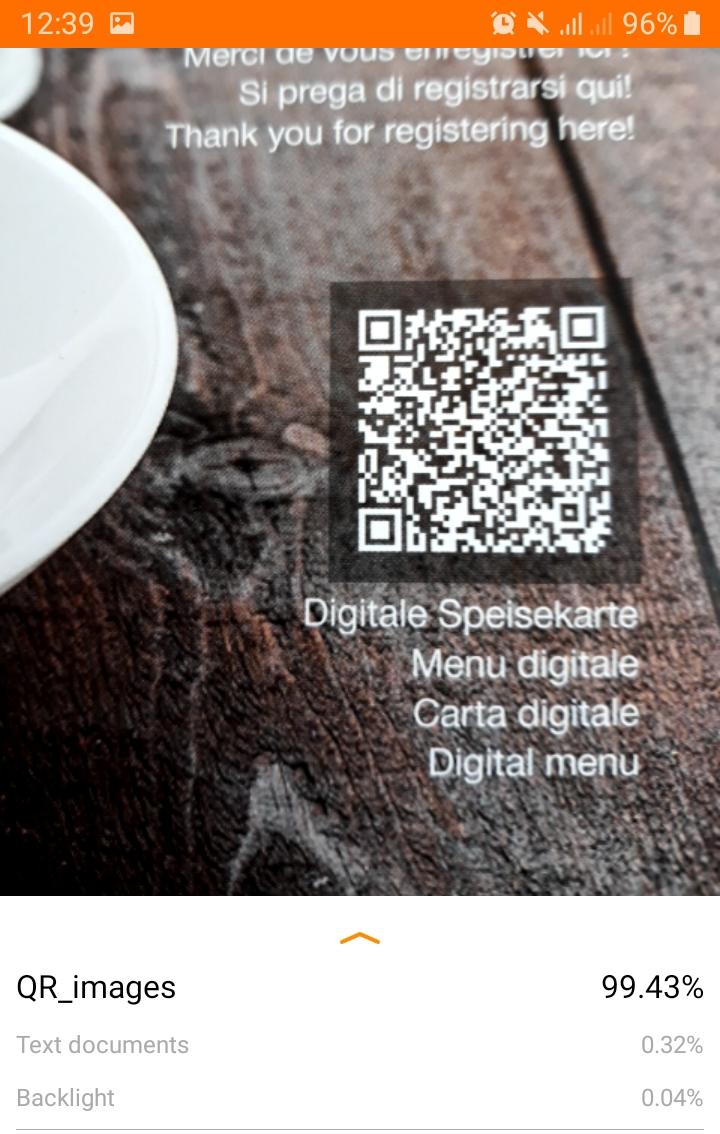}\\
   \includegraphics[width=0.32\linewidth]{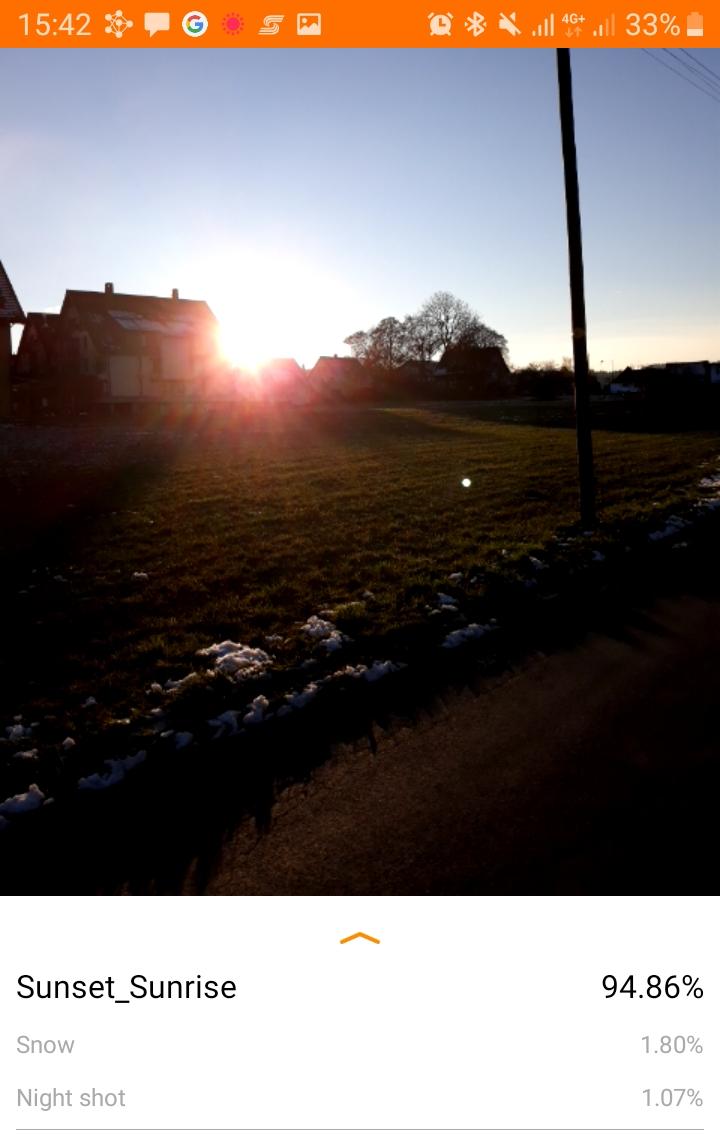}&
   \includegraphics[width=0.32\linewidth]{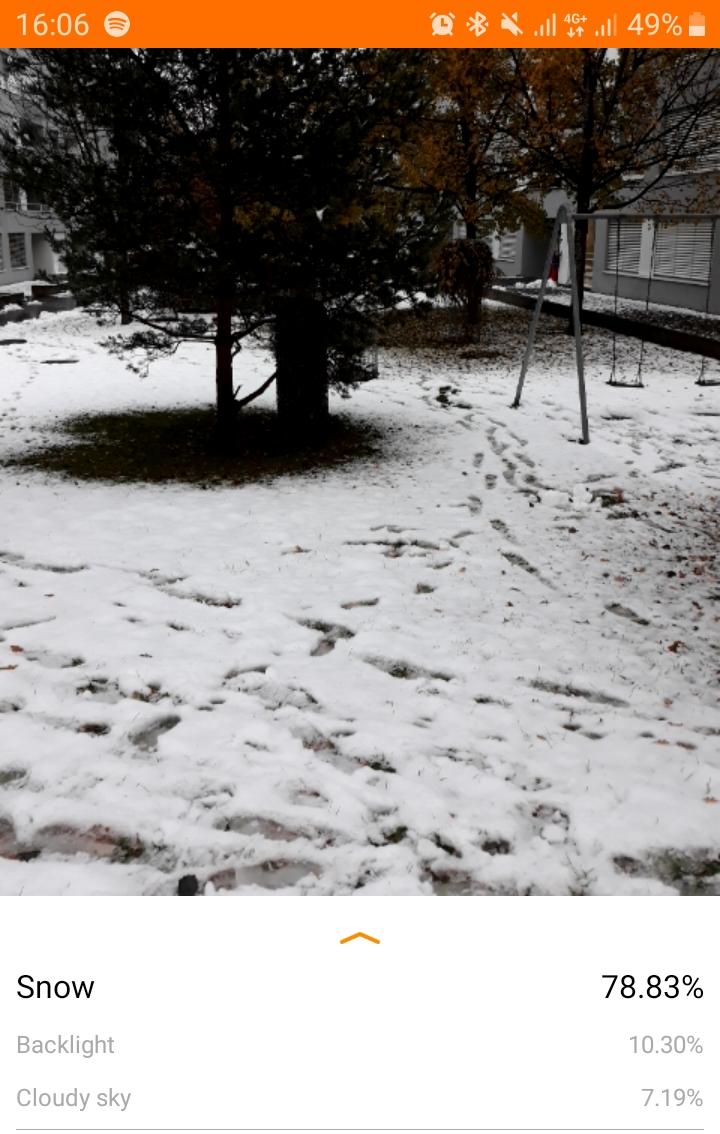}&
   \includegraphics[width=0.32\linewidth]{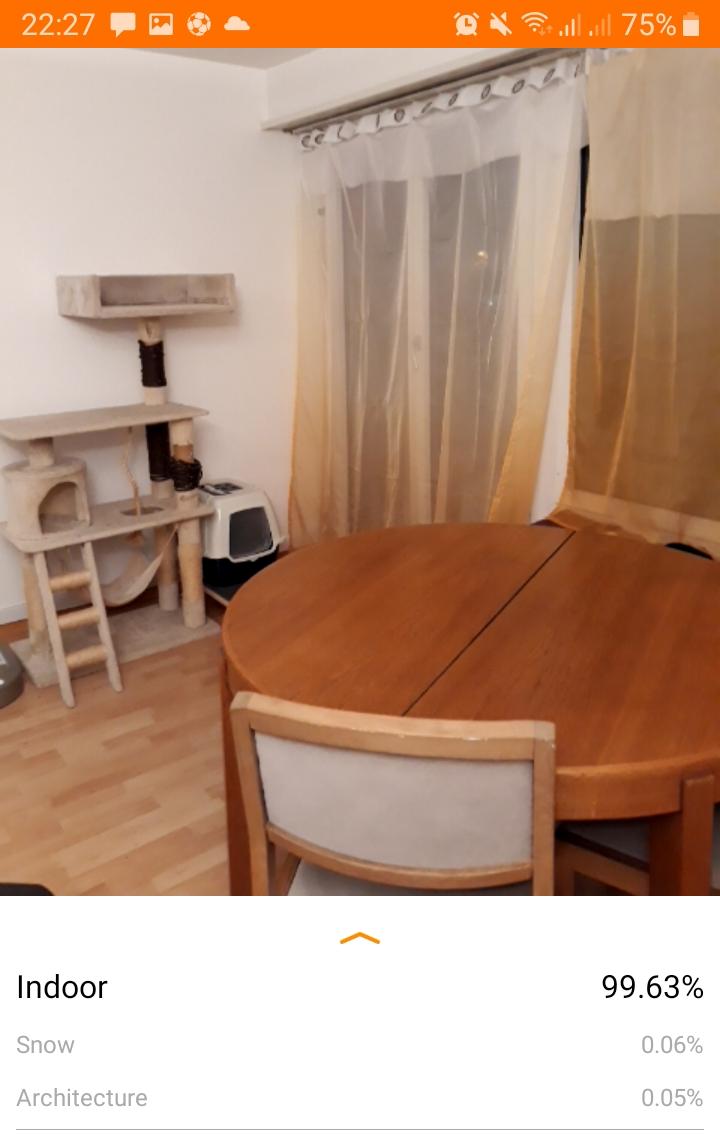}&
   \includegraphics[width=0.32\linewidth]{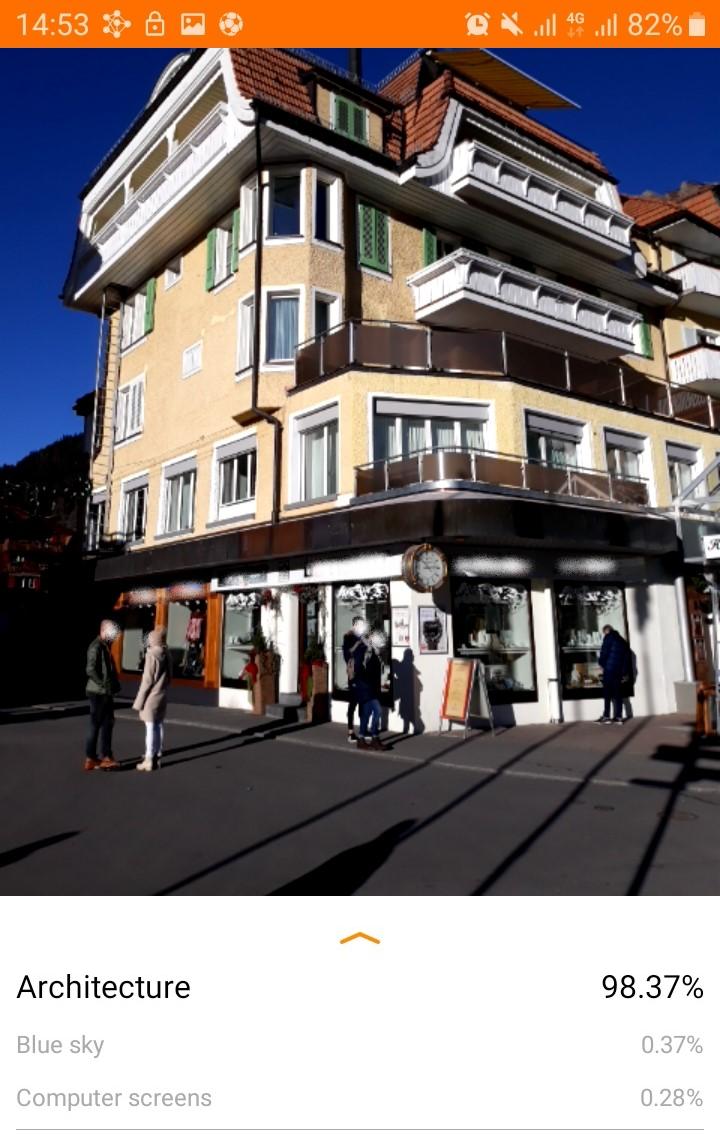}&
   \includegraphics[width=0.32\linewidth]{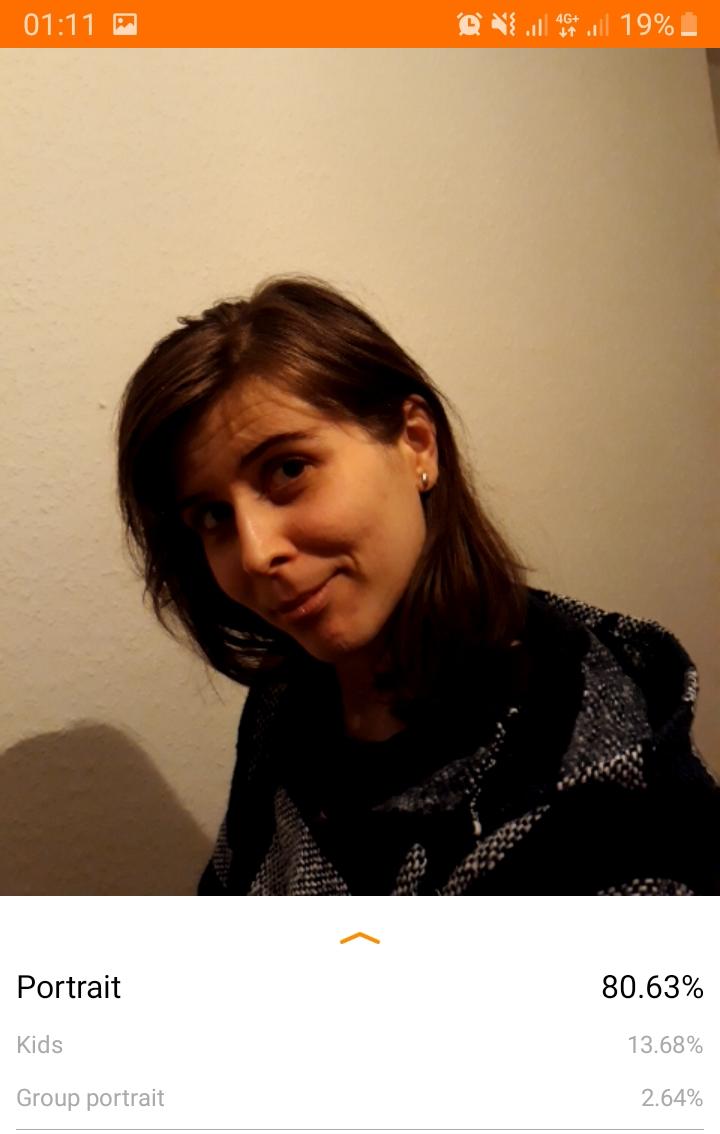}&
   \includegraphics[width=0.32\linewidth]{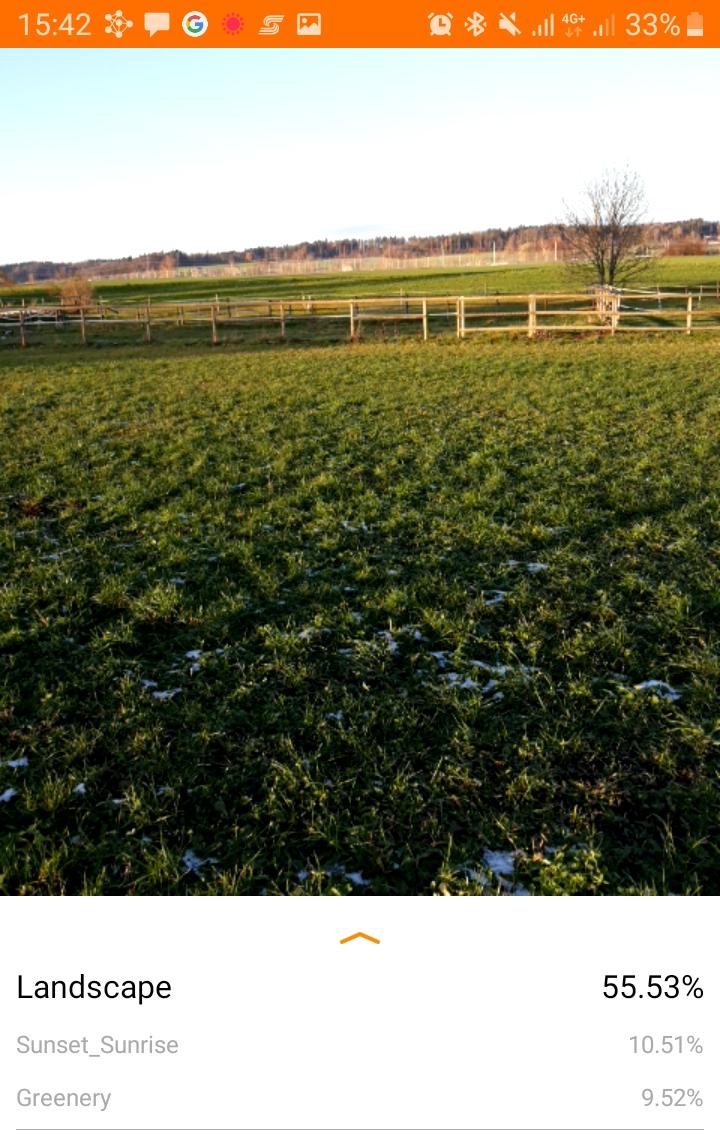}\\
   \includegraphics[width=0.32\linewidth]{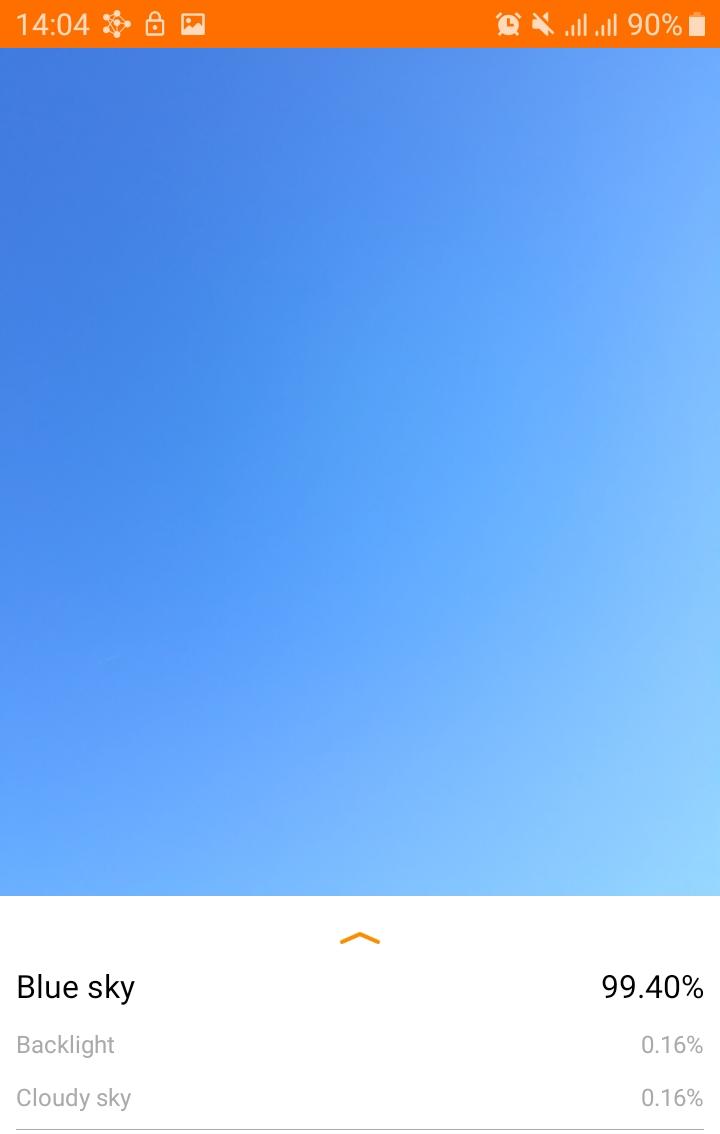}&
   \includegraphics[width=0.32\linewidth]{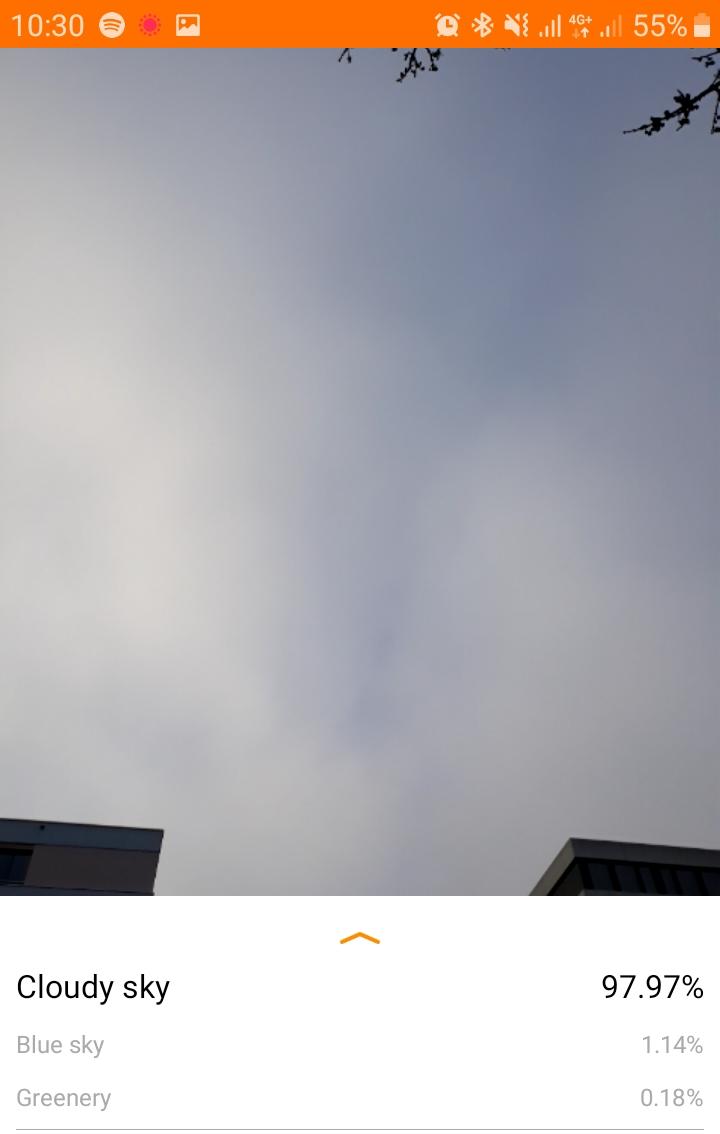}&
   \includegraphics[width=0.32\linewidth]{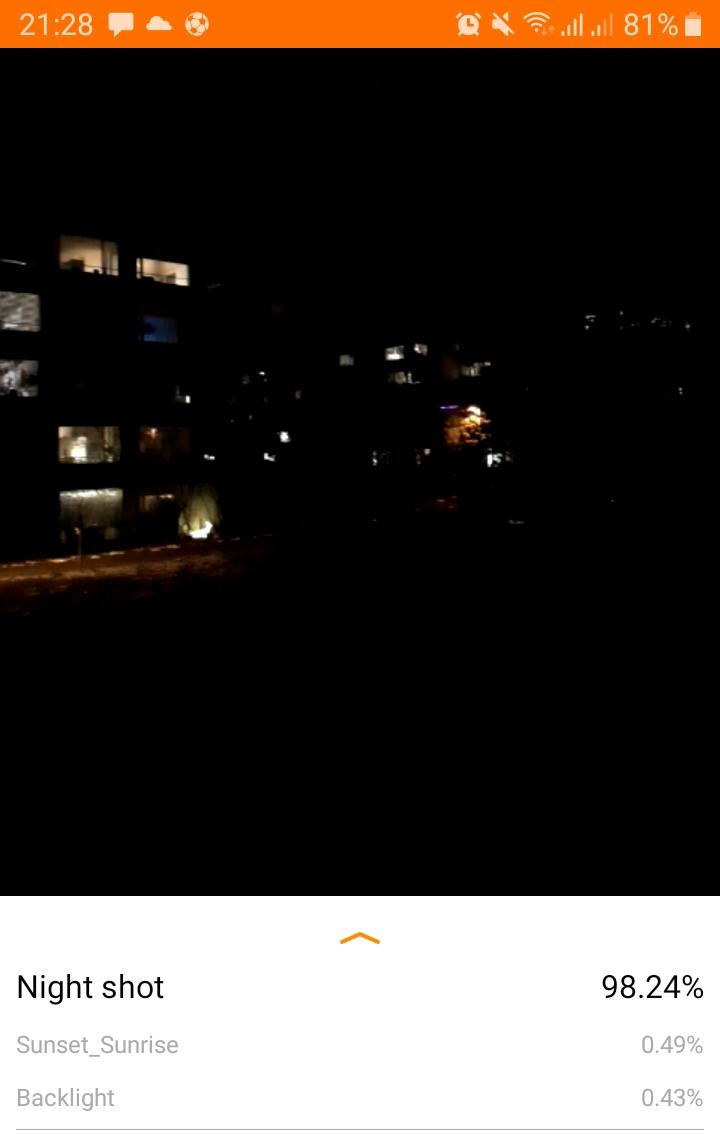}&
   \includegraphics[width=0.32\linewidth]{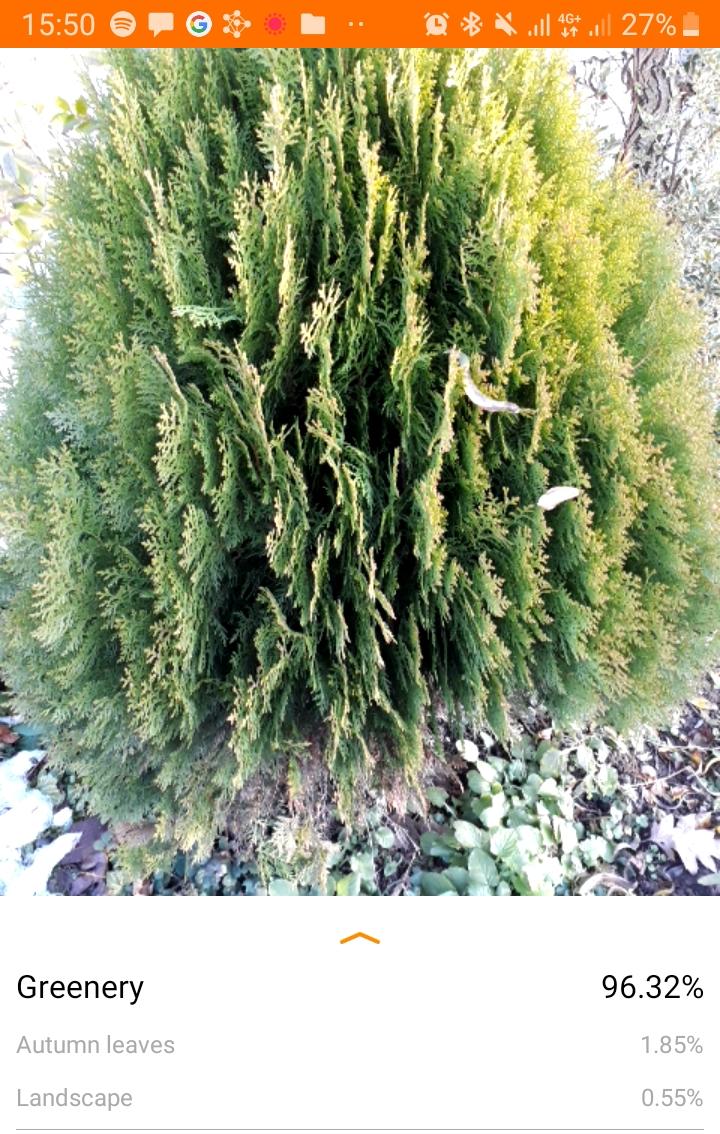}&
   \includegraphics[width=0.32\linewidth]{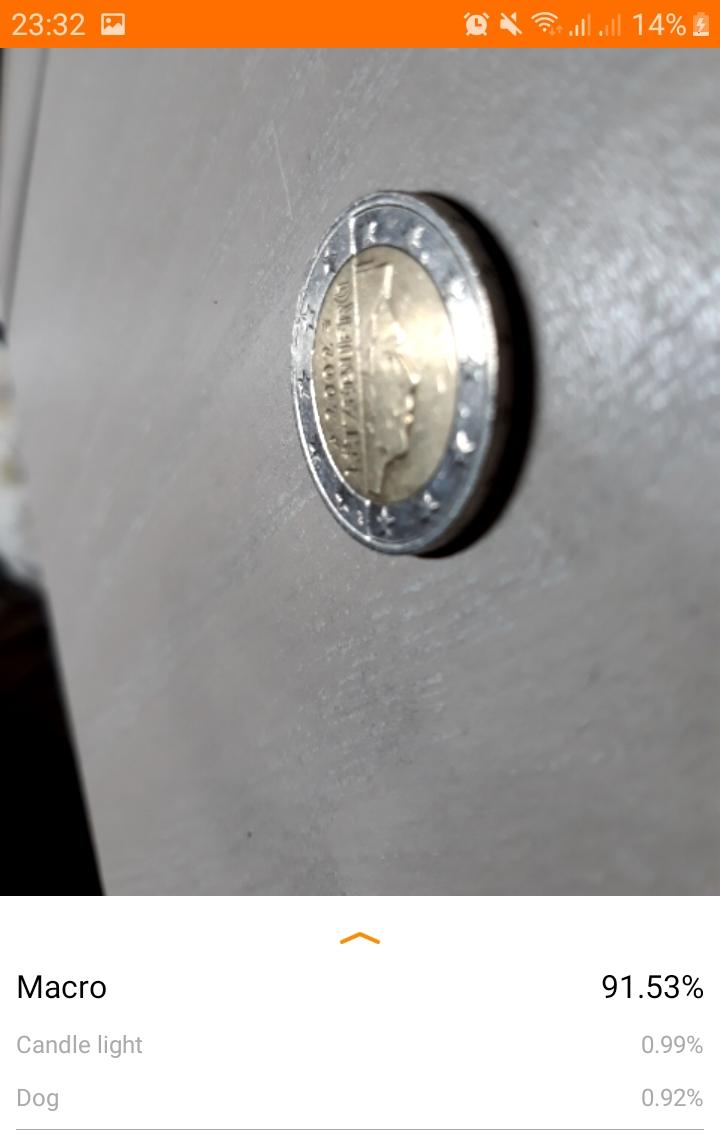}&
   \includegraphics[width=0.32\linewidth]{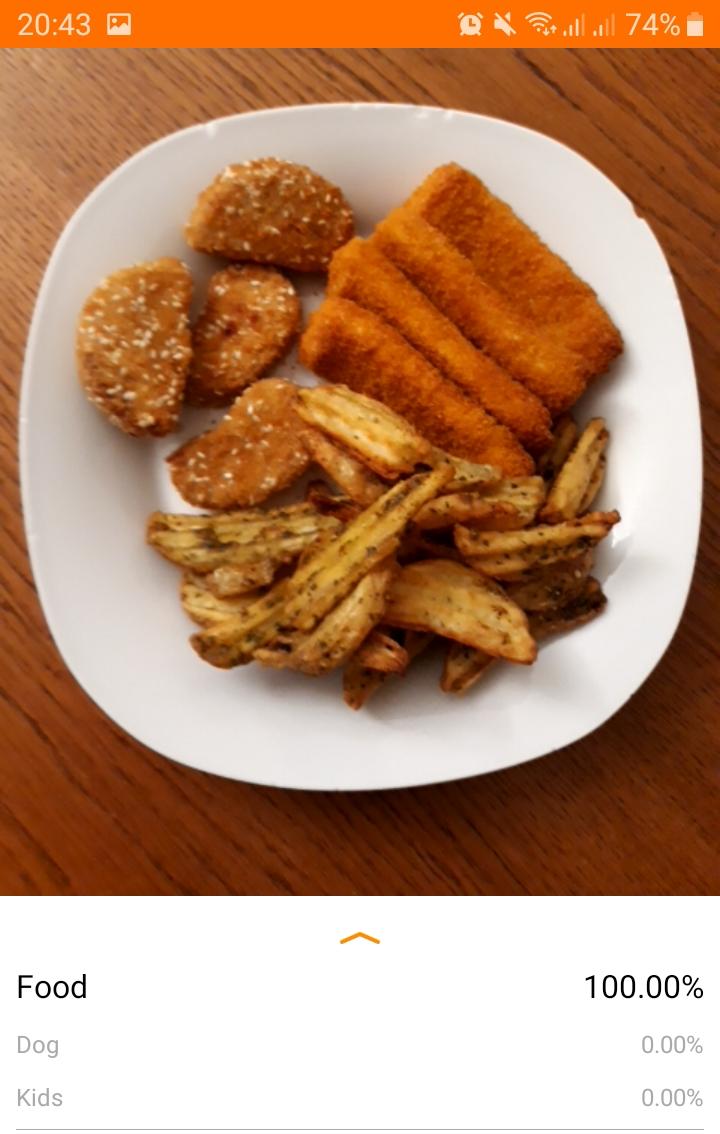}\\
\end{tabular}
}
\vspace{1.2mm}
\caption{Sample predictions obtained with the proposed MobileNet based models in-the-wild using real smartphone camera data.}
\label{fig:ideal_predictions}
\end{figure*}

\section{Experiments}
\label{sec:experiments}

\begin{figure*}[t!]
\resizebox{\linewidth}{!}
{
\begin{tabular}{cccccc}
   \includegraphics[width=0.32\linewidth]{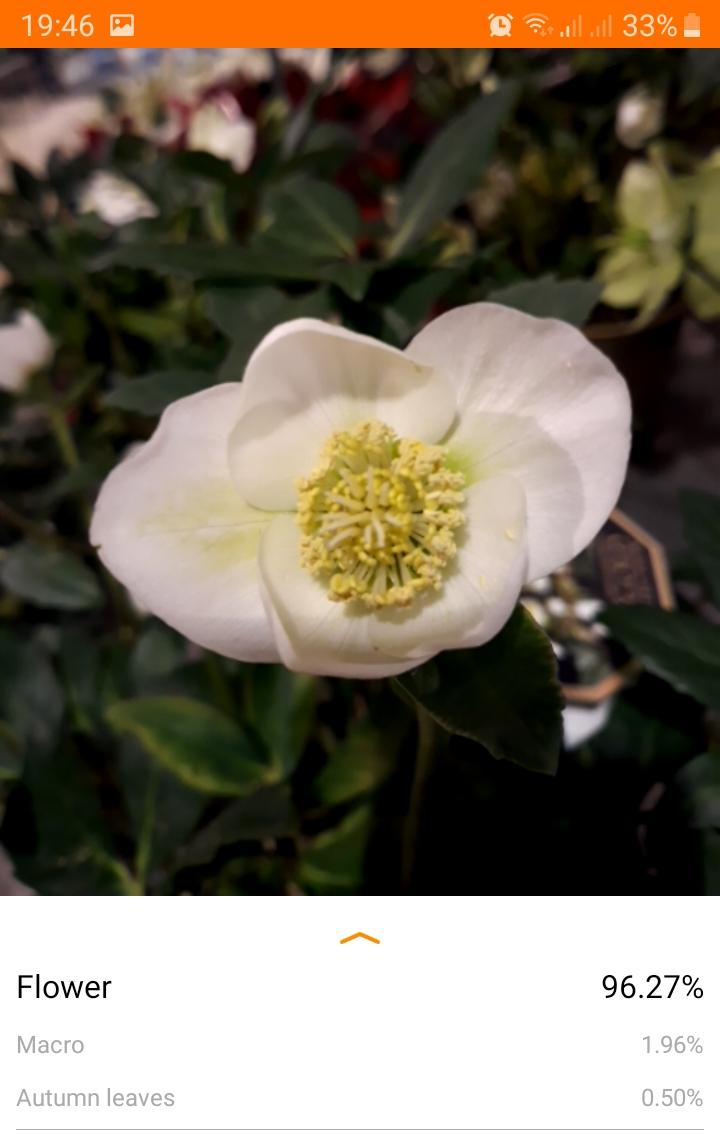}&
   \includegraphics[width=0.32\linewidth]{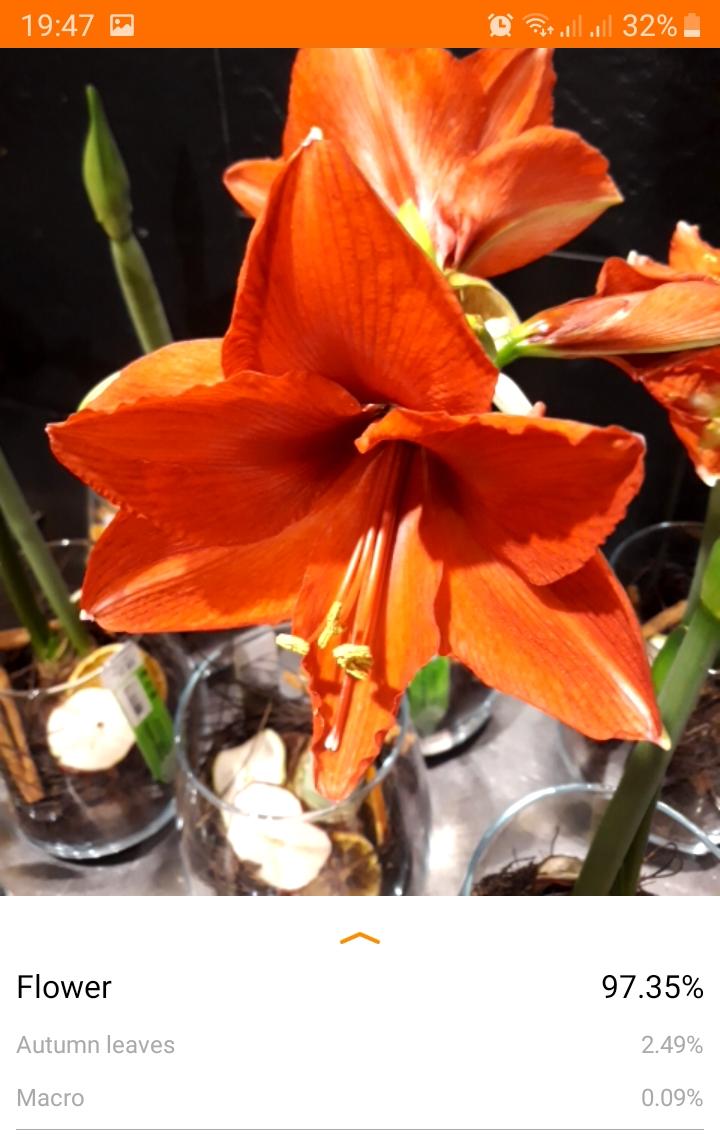}&
   \includegraphics[width=0.32\linewidth]{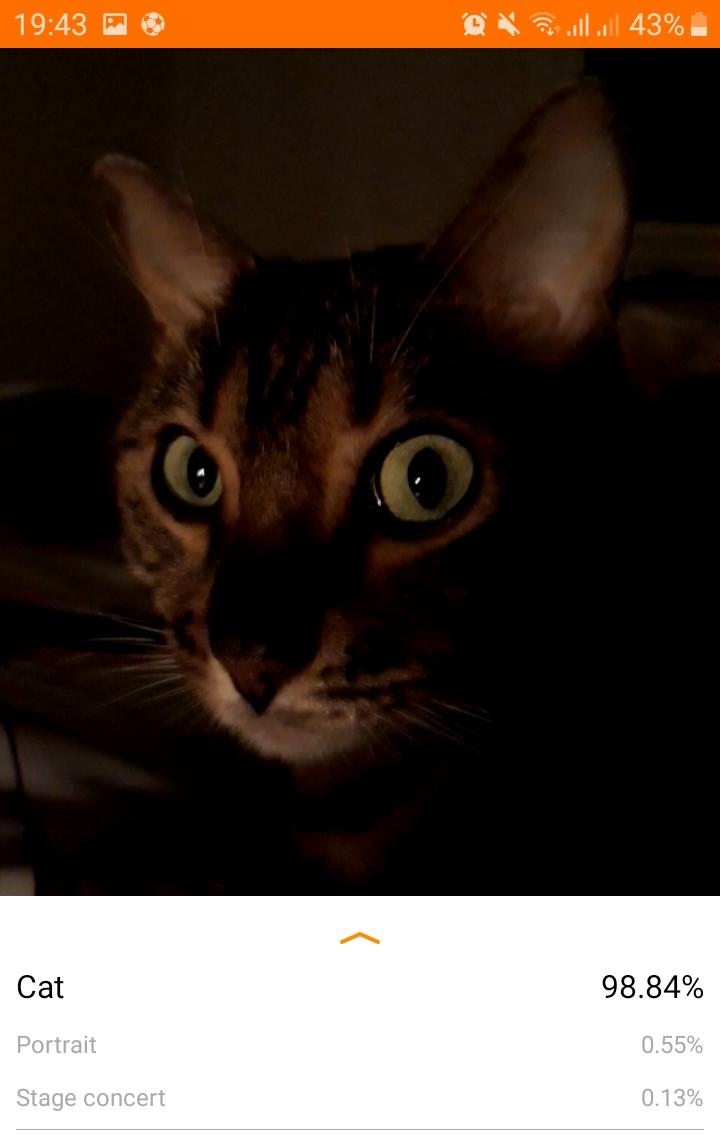}&
   \includegraphics[width=0.32\linewidth]{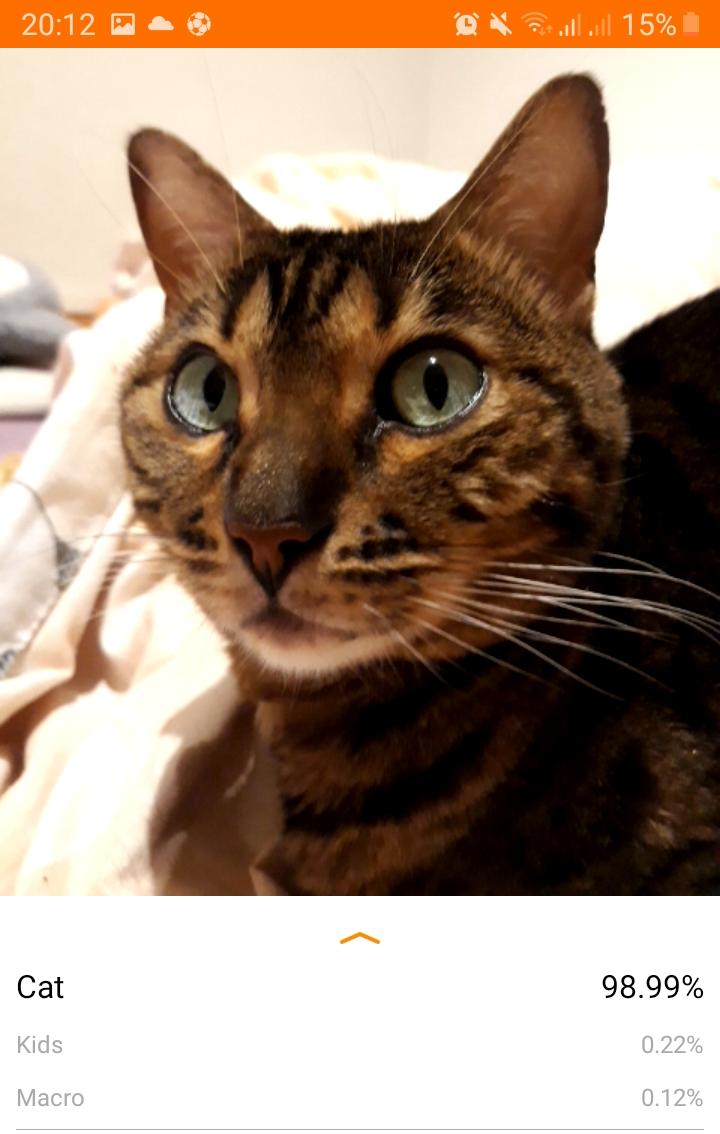}&
   \includegraphics[width=0.32\linewidth]{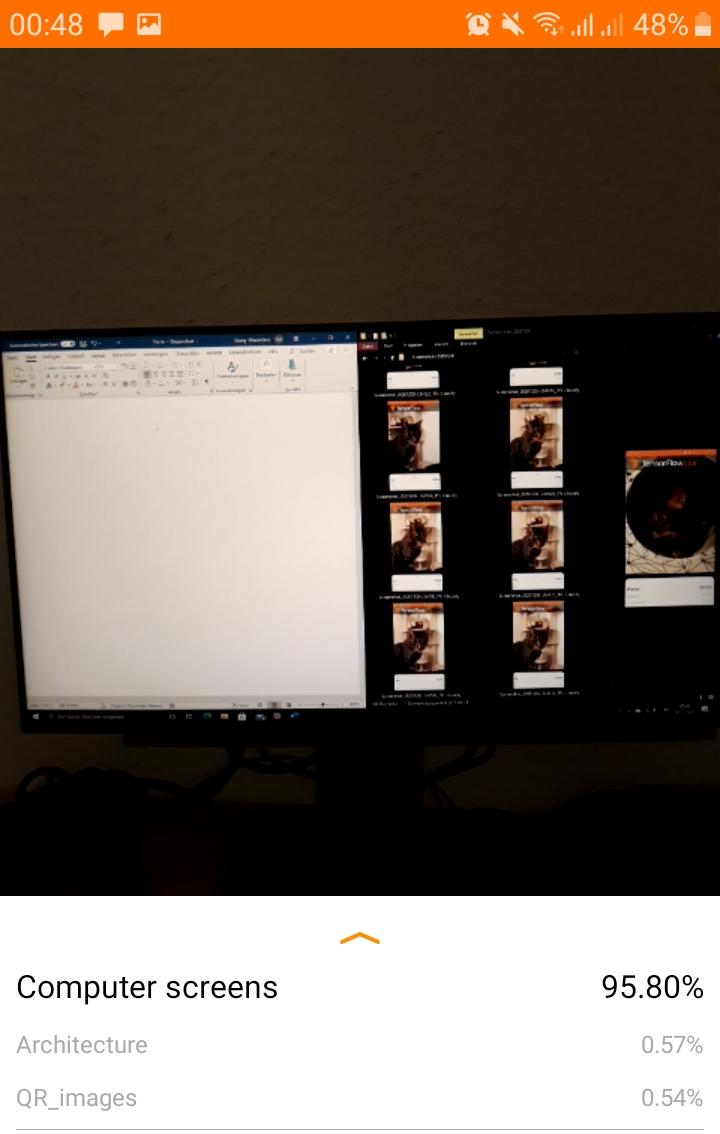}&
   \includegraphics[width=0.32\linewidth]{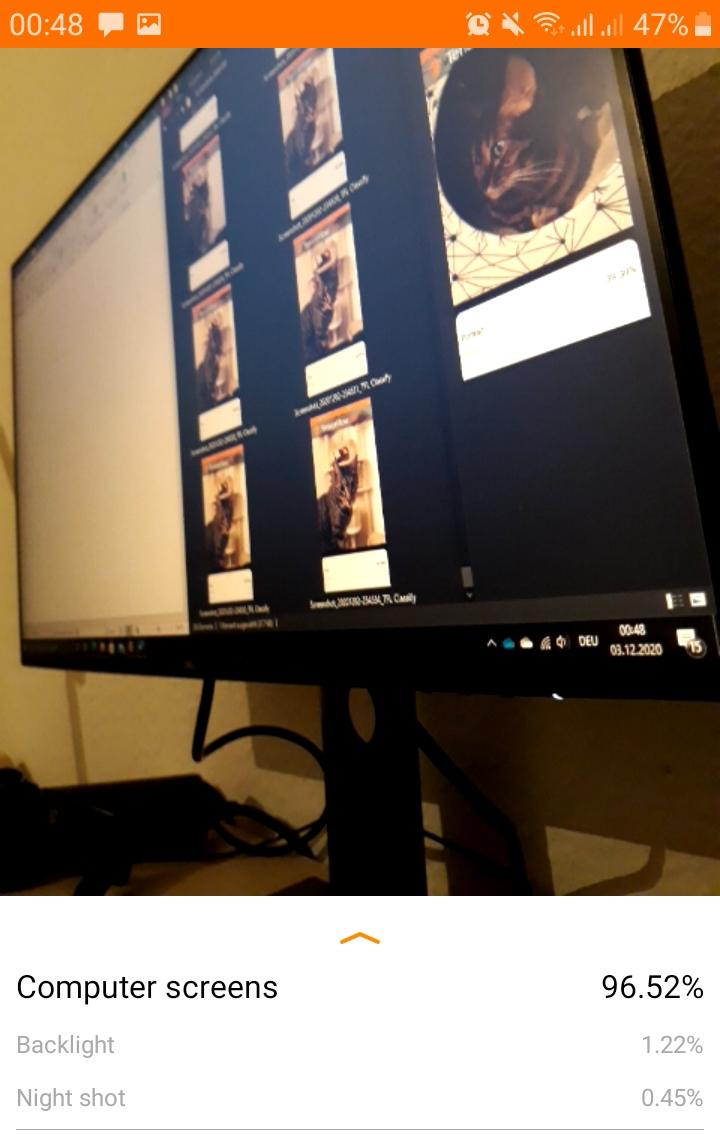}\\
\end{tabular}
}
\vspace{1.2mm}
\caption{Model predictions for different object types (left), illumination conditions (middle) and viewpoints (right).}
\label{fig:image_variations}
\end{figure*}

This section provides quantitative and qualitative results of the designed solutions as well as their runtime on several popular mobile platforms.

\subsection{Quantitative Results}

Table~\ref{tab:quant_results} presents the results obtained on the test subset of the CamSDD dataset. All models except for the one based on MobileNet-V2 are using the same fully connected feature processing block on top of them as the MobileNet-V1 model. As one can see, the first two networks were able to achieve a top-3 accuracy of more than 98\%, thus being able to identify the correct scene with a very high precision. This already suggests that the proposed setup and data works efficiently for the considered scene classification task, and the models are able to learn the underlying categorization function. The architecture based on MobileNet-V1 features achieved a top-1 accuracy of $92.67\%$ and a top-3 accuracy of $99.50\%$, outperforming all other solution by at least $0.50\%$ in the latter term. The MobileNet-V2 based network demonstrated a considerably higher top-1 accuracy of $94.17\%$ while also showing a drop of $0.83\%$ in the top-3 score, which might first seem to be counterintuitive. However, this can be explained by the fact that MobileNet-V2 features are known to be more accurate but at the same time less general than the ones produced by MobileNet-V1: while for standard scenes this results in higher predictive accuracy, these features might not be that efficient for complex and challenging conditions that the model has not seen during the training. Ideally, the best results might be achieved by combining the features and / or predictions from both models, though this is not the focus of this paper targeted at a single-backbone architecture, and can be explored in the future works. Interestingly, neither of the considered larger and allegedly more precise (in terms of the accuracy on the ImageNet) models performed good on this task, partially because of the same reason as in case of MobileNet-V2: less general features almost always result in less accurate predictions on real unseen data. Therefore, in our case we are able to get the best numerical performance with the smallest and fastest models which is ideal for a mobile-focused task.

Table~\ref{tab:quant_results} additionally reports the accuracy of the quantized MobileNet-V1/V2 based models. INT8 quantization was performed using TensorFlow's built-in post-training quantization tools~\cite{tensorflow2021postquant}. The accuracy of the MobileNet-V2 based network remained the same after applying this procedure, while the first model experienced a significant performance drop of $1.17\%$ and $0.5\%$ for top-1 and top-3 scores, respectively. Nevertheless, these results are better than the ones obtained with the other larger floating-point solutions, thus this model can be practically useful in situations when either high classification speed is needed, or for NPUs / hardware not supporting floating-point inference. The difference between the speed of the floating-point and quantized networks will be examined in the next section.

\subsection{Runtime on Mobile Devices}

To test the speed of the developed solutions on real mobile devices, we used the publicly available \textit{AI Benchmark} application~\cite{ignatov2018ai,ignatov2019ai} that allows to load any custom TensorFlow Lite model and run it on any Android device with all supported acceleration options. This tool contains the latest versions of \textit{Android NNAPI, TFLite GPU, Hexagon NN, Samsung Eden} and \textit{MediaTek Neuron} delegates, therefore supporting all current mobile platforms and providing the users with the ability to execute neural networks on smartphone NPUs, APUs, DSPs, GPUs and CPUs. To reproduce the runtime results reported in this paper, one can follow the next steps:

\begin{enumerate}
\setlength\itemsep{0mm}
\item Download AI Benchmark from the official website\footnote{\url{https://ai-benchmark.com/download}} or from the Google Play\footnote{\url{https://play.google.com/store/apps/details?id=org.benchmark.demo}} and run its standard tests.
\item After the end of the tests, enter the \textit{PRO Mode} and select the \textit{Custom Model} tab there.
\item Rename the exported TFLite model to \textit{model.tflite} and put it into the \textit{Download} folder of the device.
\item Select mode type \textit{(INT8, FP16, or FP32)}, the desired acceleration/inference options and run the model.
\end{enumerate}

\noindent These steps are also illustrated in Fig.~\ref{fig:ai_benchmark_custom}. This setup was used to test the runtime of the considered four models on 11 popular smartphone chipsets providing AI acceleration with their NPUs, DSPs and GPUs. The results of these measurements are reported in Table~\ref{tab:runtime_results}. For \mbox{MediaTek} devices, all models were accelerated on their AI Processing Units (APUs) using Android NNAPI. In case of Qualcomm chipsets, floating-point networks were accelerated with the TFLite GPU delegate demonstrating the lowest latency, while quantized networks were executed with Qualcomm's Hexagon NN TFLite delegate that performs all computations on Hexagon DSPs. On the Exynos chipsets we used either the Samsung Eden delegate or NNAPI depending on which option resulted in better runtimes, and for Huawei SoCs NNAPI was used for all four networks. Unfortunately, the Kirin 990/980 and the Snapdragon 888 chipsets were unable to run quantized TFLite models due to the lack of support for several INT8 operators, thus we had to run these networks on their CPUs with the XNNPACK delegate.

We were able to achieve real-time performance with more than 33 classified images per second on all considered platforms. Overall, the MobileNet-V2 based model turned out to be a bit faster on average than the model using MobileNet-V1 features. Quantized models have also demonstrated slightly better runtime, though the difference was not dramatic in the majority of cases, lying below 25-30\%. For MobileNet-V2 network, more than 100 FPS was obtained on six different platforms, the highest throughput was achieved on the Dimensity 1000+ (APU 3.0), Dimensity 800 (APU 3.0), Snapdragon 855 (Hexagon 690 DSP), Kirin 990 5G (Da Vinci NPU) and Snapdragon 888 (Adreno 660 GPU) SoCs, respectively. These results also demonstrate the efficiency of dedicated mobile AI processors for image classification tasks: they can achieve enormous processing rates while maintaining low power consumption. We can especially distinguish the 6-core APU found in the Dimensity 1000+ platform that has significantly outperformed all other NPUs and DSPs with more than 200 FPS for all four MobileNet models.

\subsection{In-the-wild Testing and Limitations}

\begin{figure}[t!]
\centering
{
\begin{tabular}{cc}
   \includegraphics[width=0.4\columnwidth]{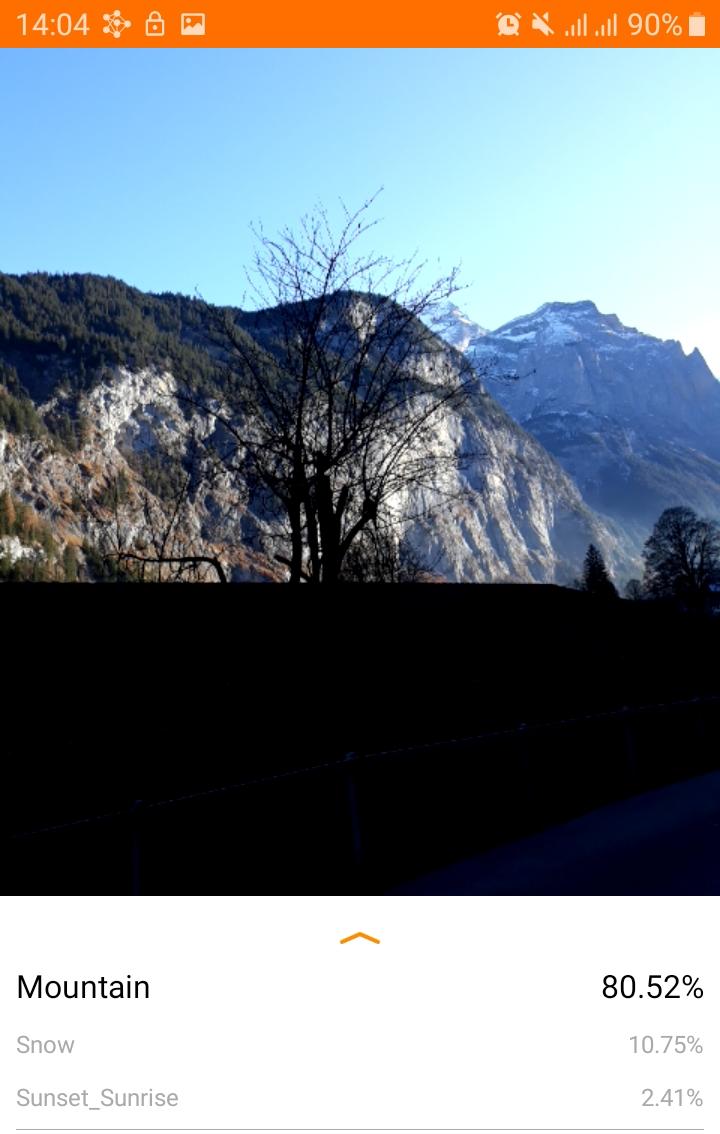}&
   \includegraphics[width=0.4\columnwidth]{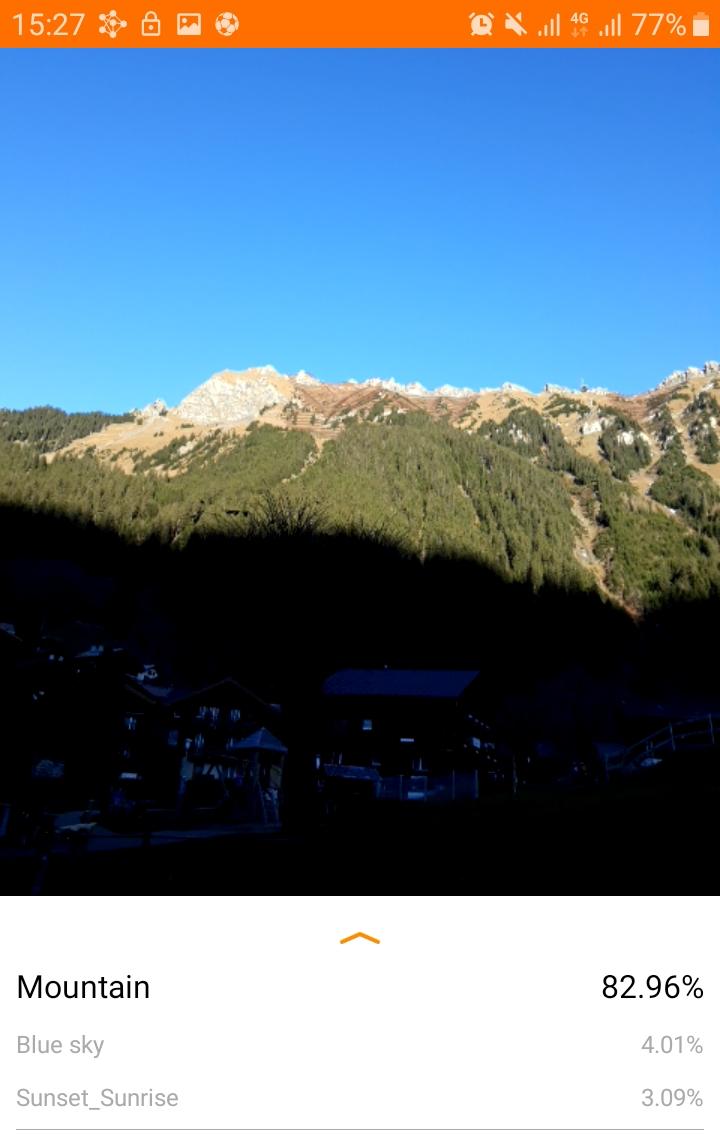}\\
   \includegraphics[width=0.4\columnwidth]{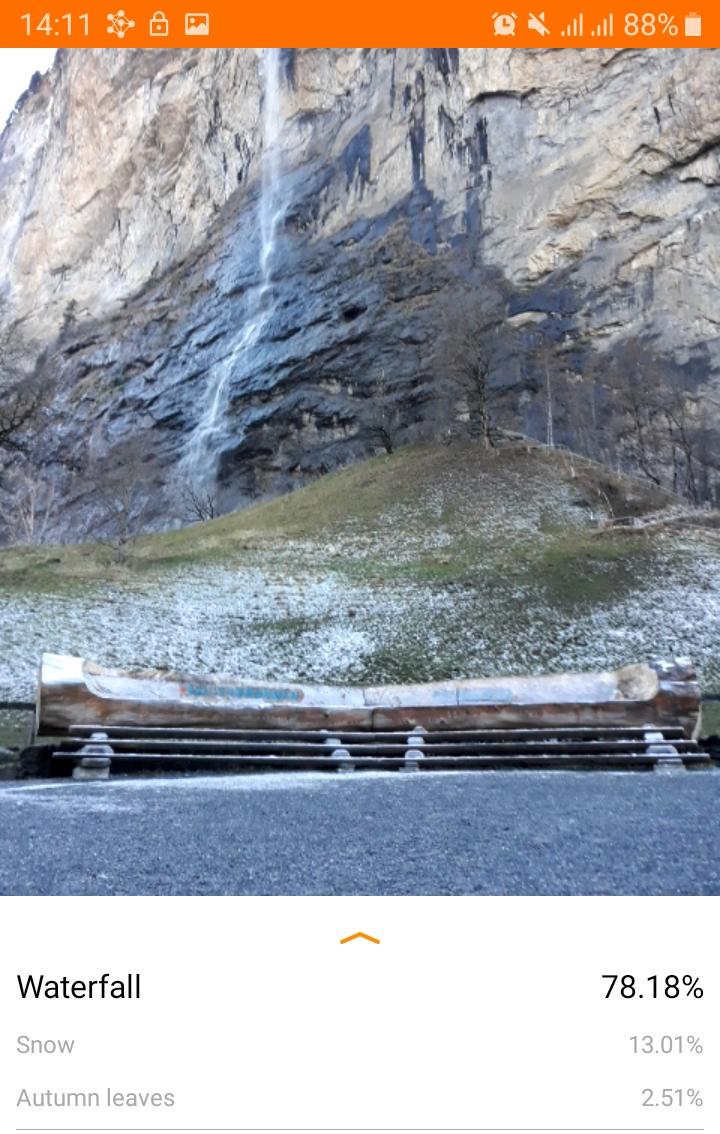}&
   \includegraphics[width=0.4\columnwidth]{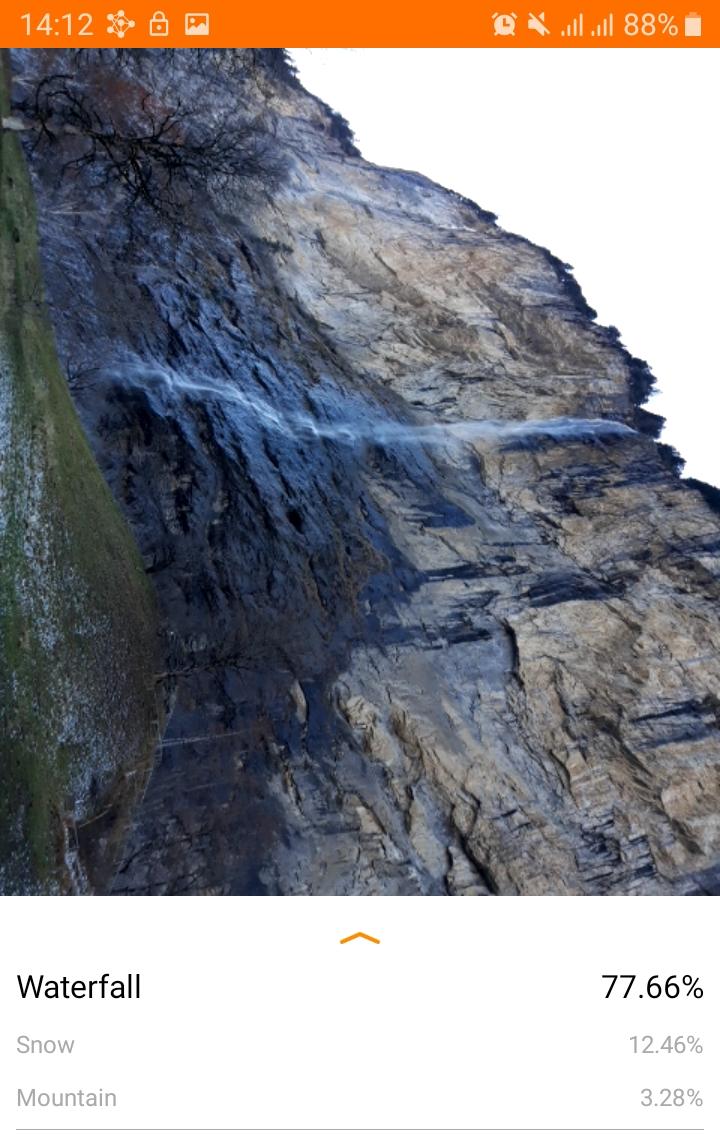}\\
\end{tabular}
}
\vspace{2mm}
\caption{Sample predictions for mountain and waterfall images.}
\label{fig:mountains_waterfall}
\vspace{0mm}
\end{figure}

\begin{figure}[t!]
\centering
{
\begin{tabular}{cc}
   \includegraphics[width=0.4\columnwidth]{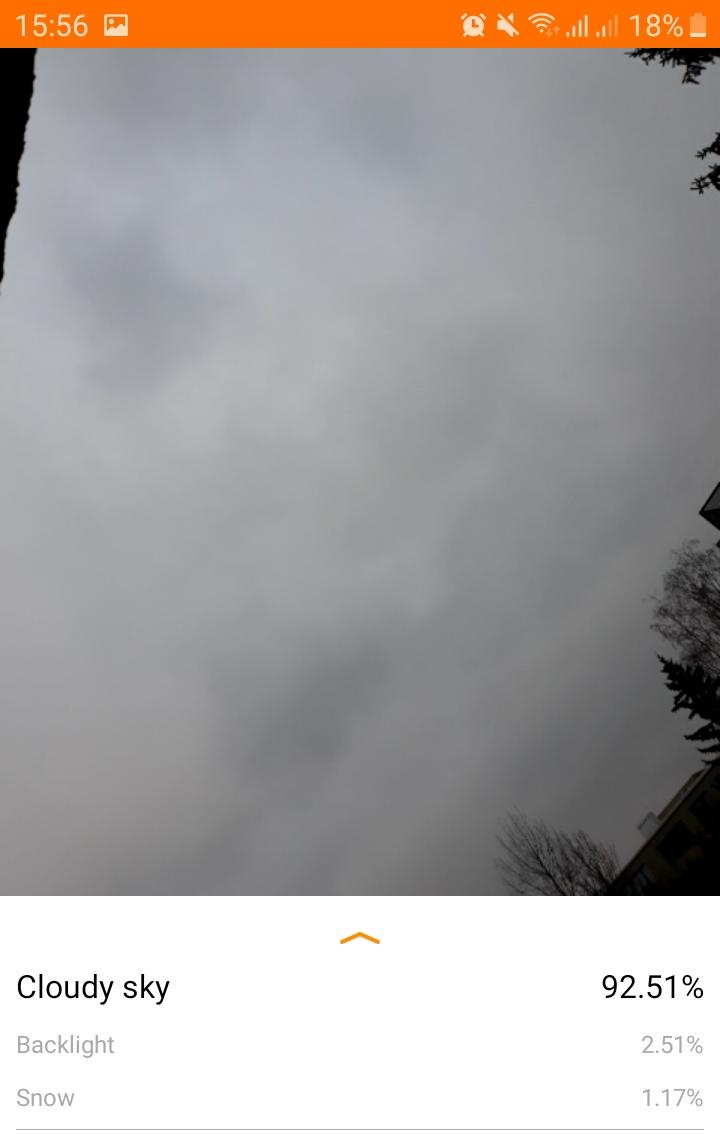}&
   \includegraphics[width=0.4\columnwidth]{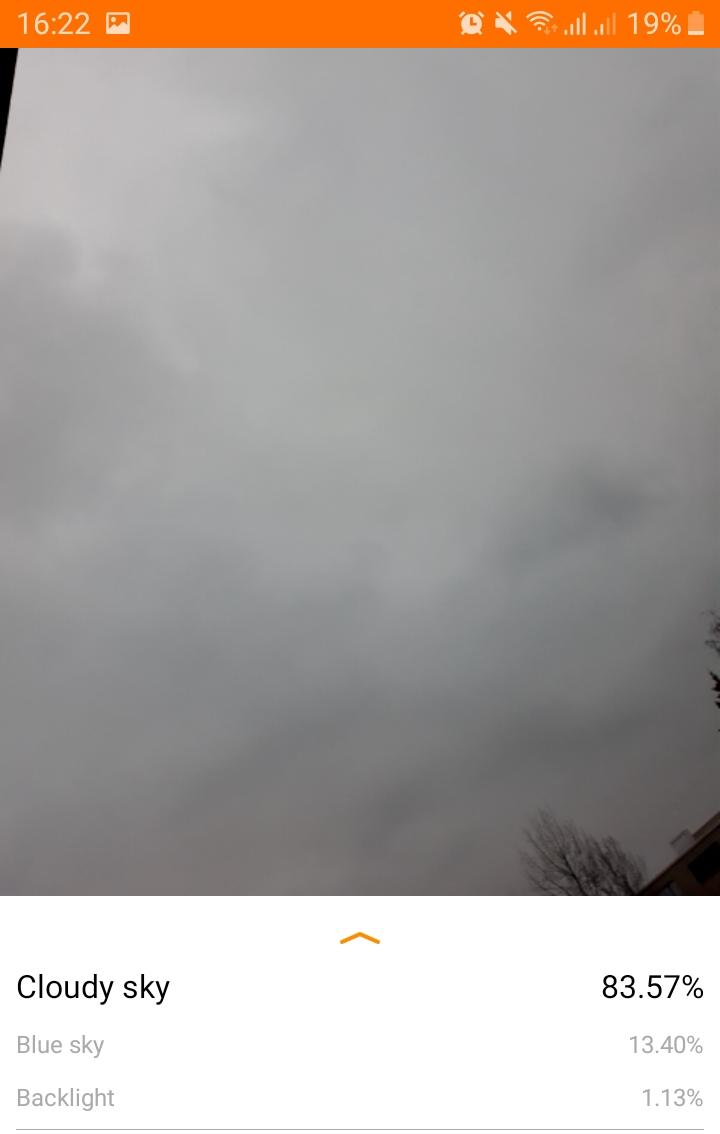}\\
   \includegraphics[width=0.4\columnwidth]{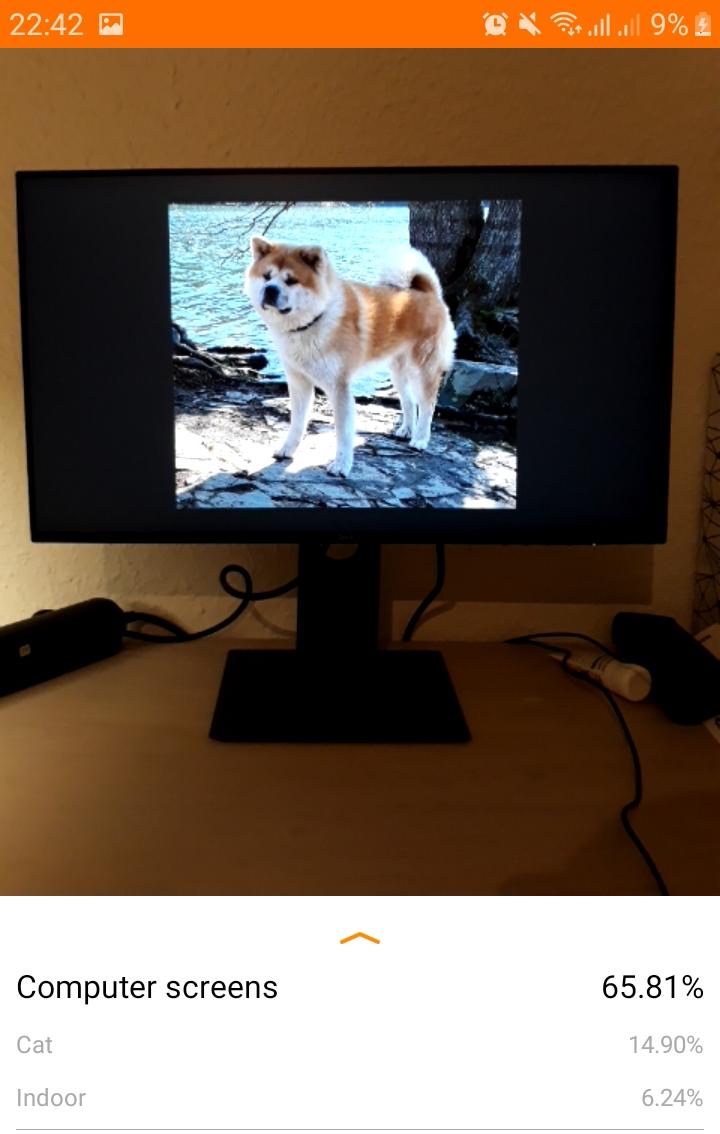}&
   \includegraphics[width=0.4\columnwidth]{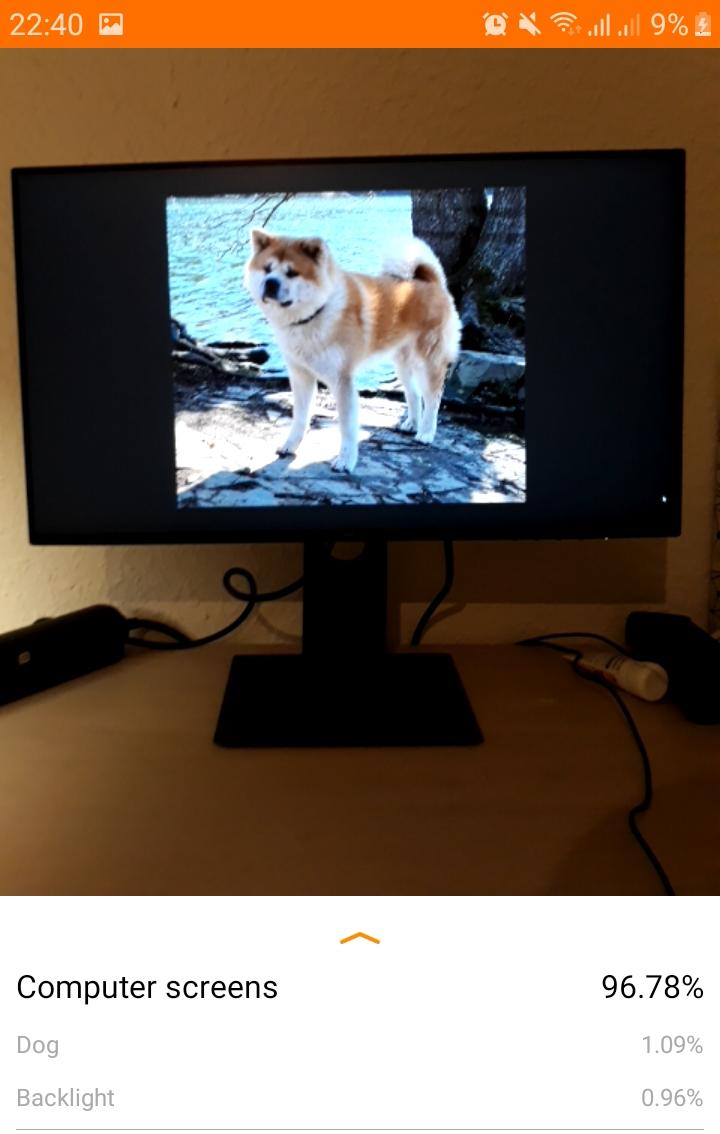}\\
\end{tabular}
}
\vspace{2mm}
\caption{The predictions of the same scene obtained using the MobileNet-V1 (left) and MobileNet-V2 (right) based models.}
\vspace{-3.2mm}
\label{fig:v1_v2_comparison}
\end{figure}

While the proposed models demonstrated high accuracy on the CamSDD dataset, their real performance on live camera data is the most important for this task. For this, we developed an Android application that is using the obtained TensorFlow Lite models to perform real-time classification of the image frames coming from camera stream. The general design of the application is similar to~\cite{tflite2021camerademo}. Two popular smartphones were used for testing: the Samsung Galaxy J5 and the Samsung Galaxy S9. We checked the predictions of the developed models on hundreds of different scenery, and present in this section the most important observations. Since the Samsung Galaxy J5 is equipped with a low-end camera whose quality is considerably worse compared to the majority of modern smartphones, including the S9 one, this was our main target device as the conditions in this case are the most challenging. Therefore, if not stated otherwise, the presented screenshots refer to the Galaxy J5.

The overall accuracy of the presented solution is very satisfactory when testing it on real camera data. As one can see in Fig.~\ref{fig:ideal_predictions}, it is able to correctly predict the standard scene categories such as \textit{Architecture}, \textit{Flower}, \textit{Portrait}, \textit{Candlelight}, \etc, with a very high confidence. In general, we obtained robust results when facing the following challenges. First, the model was robust towards intra-class variation, \ie, the variation between the images belonging to the same class. For instance, in Fig.~\ref{fig:image_variations} one can see correct predictions for two flower types that greatly vary in shape and color. Secondly, it can handle large illumination changes (Fig.~\ref{fig:image_variations}, middle) and was also robust towards view-point variations (Fig.~\ref{fig:image_variations}, right): as can be seen on these images, the cat and the screen were detected flawlessly regardless of the camera position and lighting. Furthermore, under normal illumination conditions we were able to get correct predictions for the majority of complex classes like \textit{Waterfall} or \textit{Mountain} that contain many elements from other categories such as blue / cloudy sky, snow, lake and / or greenery. For instance, in Fig.~\ref{fig:mountains_waterfall} one can see the waterfall flowing on the slope of a hill, and the image itself has many similarities to the class \textit{Mountain}. This makes it particularly difficult to make correct predictions. However, our model was able to do so as we trained it with a variety of complex scenery, \eg, for the above class we used images containing different weather conditions, mountains with and without snow as well as photos with and without lakes, greenery, \etc.

\begin{figure}[t!]
\centering
{
\begin{tabular}{cc}
   \includegraphics[width=0.39\columnwidth]{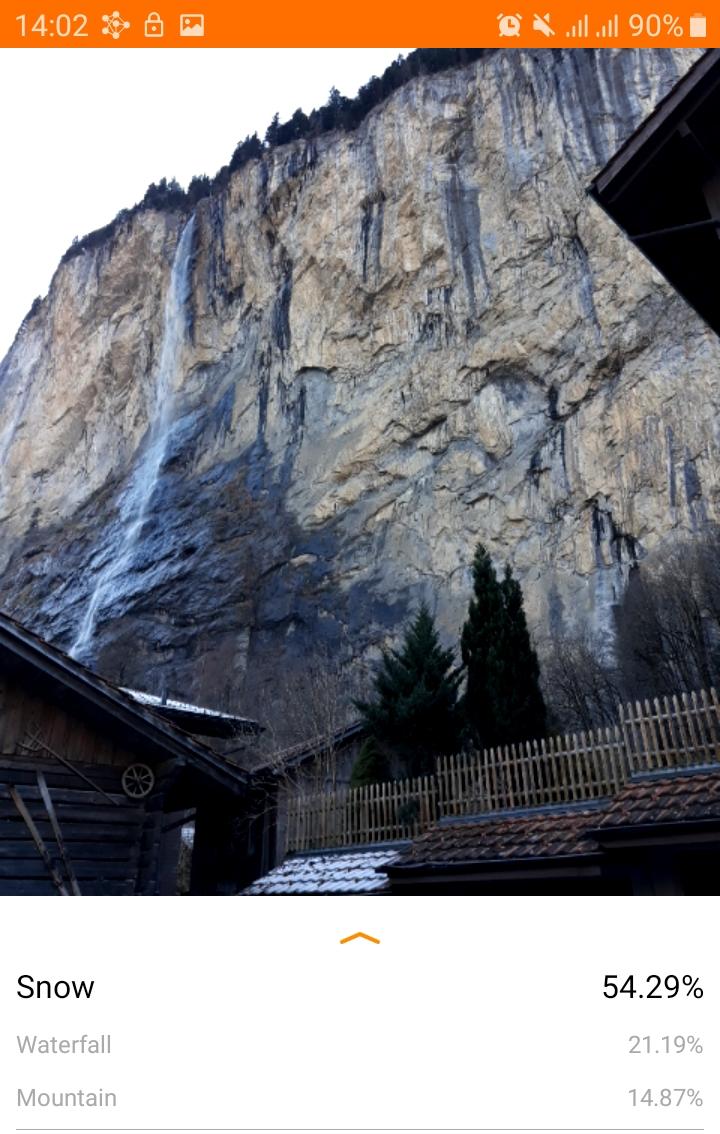}
   \includegraphics[width=0.39\columnwidth]{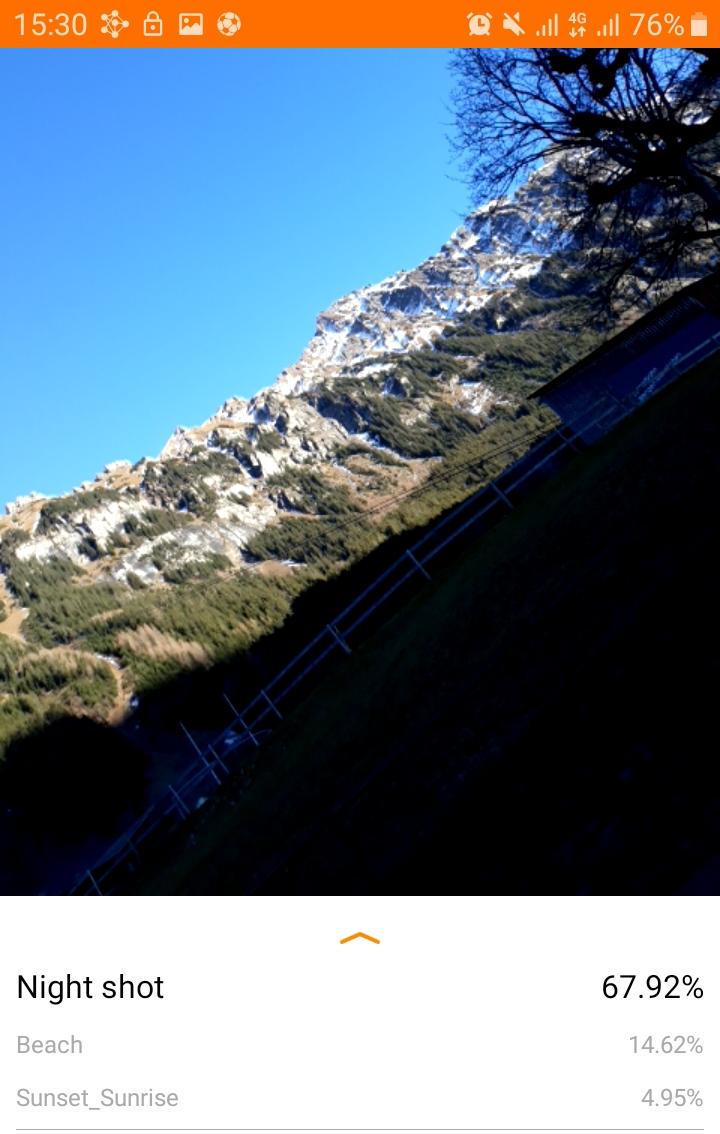}\\
\end{tabular}
}
\vspace{2mm}
\caption{Incorrect predictions for classes \textit{Mountain} and \textit{Waterfall} for images with over- and under-exposed regions.}
\label{fig:overexposed_images}
\end{figure}

Though we did not observe any major issues under good lighting conditions, some problems might appear when photos have large over- or under-exposed regions. Fig.~\ref{fig:overexposed_images} demonstrates the classification results obtained on the image with an over-exposed sky area: instead of being blue, the top left corner of the photo is completely white since the Galaxy J7 camera cannot handle HDR scenes due to the limited sensor bit-width. Though the model was still able to recognize waterfall in this case, this was only the second top prediction, and the general object class was detected as \textit{Snow}. An opposite example is shown on the right photo: as half of the image was almost completely dark, the network suggested that this is the \textit{Night Shot} scene. In general, the standard ambient light installed nowadays in any smartphone can be used to deal with this problem. Another possible solution would be a control loop that is based on the selected scene. For example, if the \textit{Night Shot} scene is predicted, the camera adjusts its ISO level to brighten up the image, and thus a better prediction could be made.

\begin{figure}[t!]
\centering
{
\begin{tabular}{cc}
   \includegraphics[width=0.4\columnwidth]{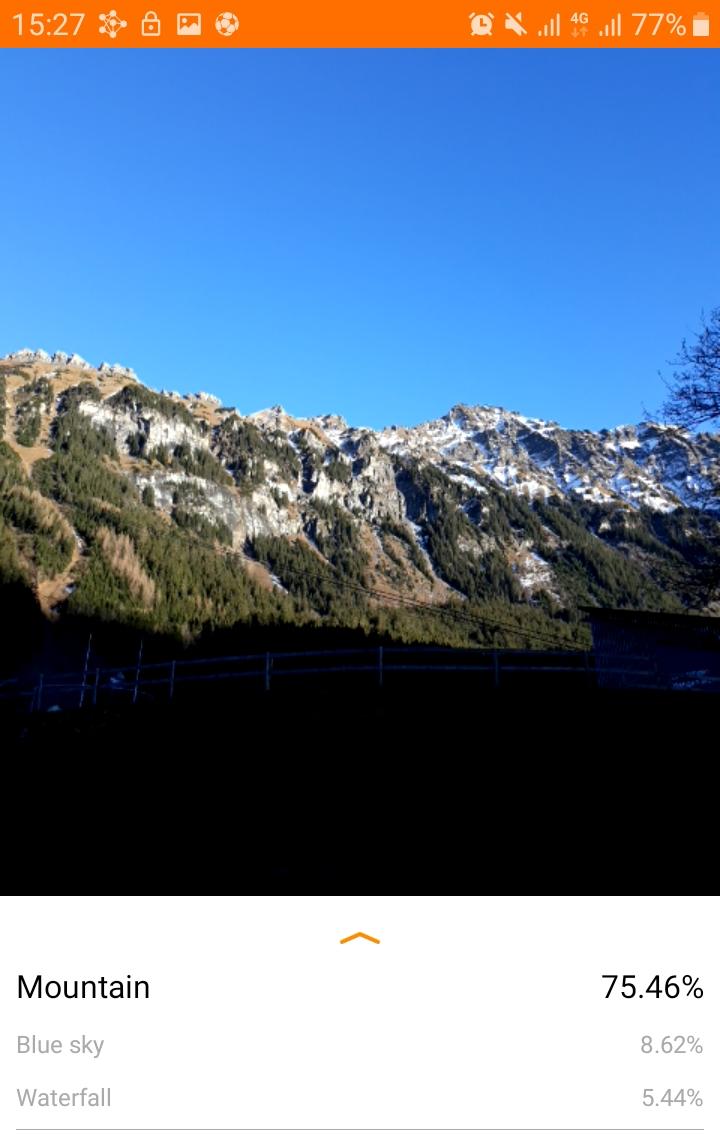}&
   \includegraphics[width=0.4\columnwidth]{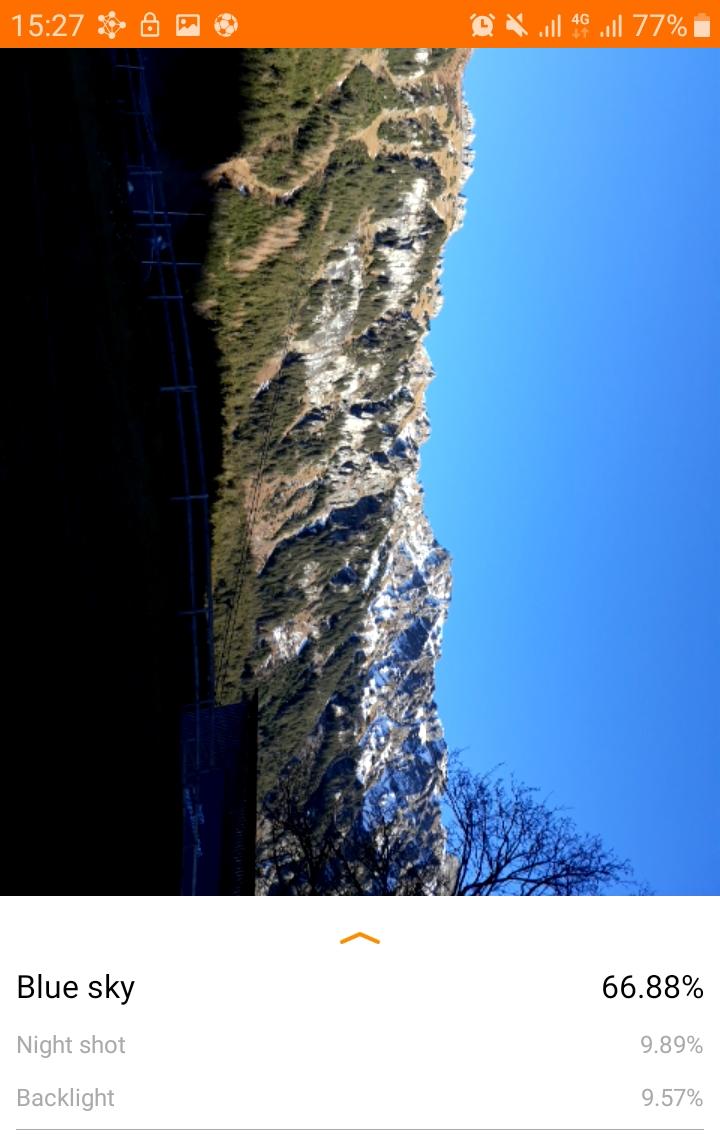}\\
\end{tabular}
}
\vspace{2mm}
\caption{Model predictions for the same mountain scene in portrait (left) and landscape (right) modes.}
\label{fig:landscape_mode_images}
\end{figure}

Two other minor problems are related to our camera app implementation. As we do not rotate the image based on gyroscope data, its position is not correct when the smartphone is in landscape mode, and thus the predictions might also be distorted as shown in Fig.~\ref{fig:landscape_mode_images}. Finally, when pointing the camera at scenery or objects that are not present in our training set, the resulting probabilities for all classes are close to zero, and thus the output is almost random. This problem can be easily fixed by adding a threshold for the probabilities obtained before the \textit{Softmax} layer: no predictions are returned if this threshold is not reached for any scene category.

\begin{figure}[b!]
\centering
{
\begin{tabular}{cc}
   \includegraphics[width=0.4\columnwidth]{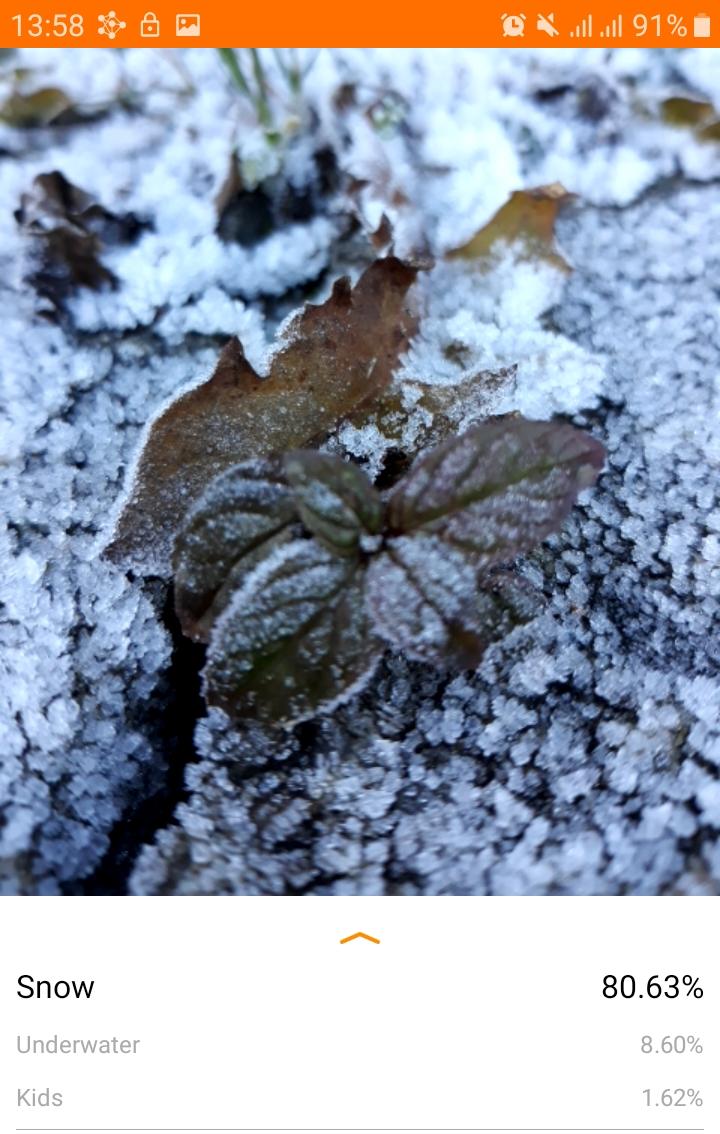}&
   \includegraphics[width=0.4\columnwidth]{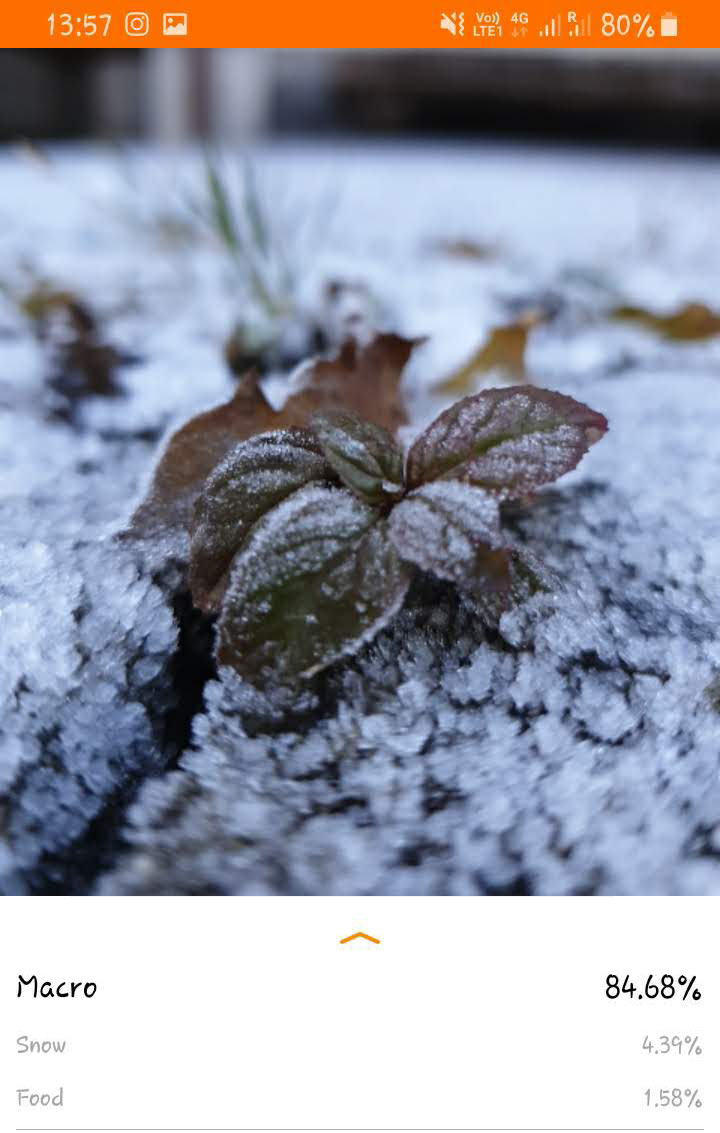}\\
\end{tabular}
}
\vspace{2mm}
\caption{Model predictions for the same \textit{Macro} scene obtained on the Samsung Galaxy J5 (left) and the Samsung Galaxy S9 (right) smartphones.}
\label{fig:macro}
\end{figure}

During our field testing we used both the MobileNet-V1 and MobileNet-V2 based models. Overall, their predictions are very close for the majority of scenes. The biggest difference between them is that the latter network produces slightly more accurate results for standard object categories such as \textit{Dog}, \textit{Screen}, \textit{Flower}, \etc., while the MobileNet-V1 is able to identify more challenging scenery like~\textit{Cloudy Sky} a bit more precisely, which aligns well with our previous observations. Otherwise, one can select one of these two models solely based on the ops / layer support, runtime and size requirements.

Lastly, the camera quality might also impact the accuracy of the obtained predictions. For instance, when trying to capture close-up images, we could not always achieve good results with the Galaxy J5. On the other hand, the Galaxy S9 performed very well as shown in Fig.~\ref{fig:macro}: it can shoot photos at closer distances and has large aperture optics resulting in greatly improved image quality compared to the Galaxy J5. Therefore, the model also performed better on the Galaxy S9 device.

\subsection{MAI 2021 Camera Scene Detection Challenge}

The considered CamSDD dataset was also used in the \textit{MAI 2021 Real-Time Camera Scene Detection Challenge}, where the goal was to develop fast and accurate quantized scene classification models for mobile devices. A detailed description of the solutions obtained in this challenge is provided in~\cite{ignatov2021fastSceneDetection}. This competition was a part of a larger Mobile AI 2021 Workshop\footnote{\url{https://ai-benchmark.com/workshops/mai/2021/}} targeted at efficient models for different mobile-related tasks such as learned smartphone ISP on mobile NPUs~\cite{ignatov2021learned}, real image denoising on mobile GPUs~\cite{ignatov2021fastDenoising}, quantized image super-resolution on Edge SoC NPUs~\cite{ignatov2021real}, real-time video super-resolution on mobile GPUs~\cite{romero2021real}, and fast single-image depth estimation on mobile devices~\cite{ignatov2021fastDepth}.

\section{Conclusion}
\label{sec:conclusion}

This paper defines the problem of efficient camera scene detection for mobile devices with deep learning. We proposed a novel large-scale CamSDD dataset for this task that is composed of 30 most vital scene categories for mobile cameras. An efficient MobileNet-based solution was developed for this problem that demonstrated a top-1/top-3 accuracy of more than 94\% and 98\%, respectively, and achieved more than 200 FPS on the latest mobile NPUs. A thorough in-the-wild testing of the proposed solution revealed its high performance and robustness to various challenging scenes, shooting conditions and environments. Finally, we made the dataset and the designed models publicly available to establish an efficient baseline solution for this task. The problem of accurate camera scene detection will also be addressed in the next Mobile AI challenges to further boost the precision and efficiency of the scene classification models.

{\small
\bibliographystyle{ieee_fullname}

}

\end{document}